\documentclass[aip,reprint]{revtex4-1}

\usepackage{graphicx}% Include figure files
\usepackage{amsmath}
\usepackage{ccaption}
\usepackage{url}
\usepackage{color}

\captionnamefont{\small\normalfont}
\captiontitlefont{\small\normalfont} \captionstyle{\flushleftright}
\captiondelim{.} \normalcaption{} \captionwidth{\linewidth}

\begin{document}

\title{QCAD Simulation and Optimization of Semiconductor Quantum Dots}

\author{X. Gao}
\email[]{xngao@sandia.gov}
\author{E. Nielsen}
\author{R. P. Muller}
\author{R. W. Young}
\author{A. G. Salinger}
\author{N. C. Bishop}
\author{M. Lilly}
\author{M. S. Carroll}
\affiliation{Sandia National Laboratories, 1515 Eubank SE, Albuquerque, NM 87123, USA}

\date{\today}

\begin{abstract}
\noindent We present the Quantum Computer Aided Design (QCAD) simulator that targets modeling multi-dimensional quantum devices, particularly silicon multi-quantum dots (QDs) developed for quantum bits (qubits). This finite-element simulator has three differentiating features: (i) its core contains nonlinear Poisson, effective mass Schrodinger, and Configuration Interaction solvers that have massively parallel capability for high simulation throughput, and can be run individually or combined self-consistently for 1D/2D/3D quantum devices; (ii) the core solvers show superior convergence even at near-zero-Kelvin temperatures, which is critical for modeling quantum computing devices; and (iii) it interfaces directly with the full-featured optimization engine Dakota. In this work, we describe the capabilities and implementation of the QCAD simulation tool, and show how it can be used to both analyze existing experimental QD devices through capacitance calculations, and aid in the design of few-electron multi-QDs. In particular, we observe that computed capacitances are in rough agreement with experiment, and that quantum confinement increases capacitance when the number of electrons is fixed in a quantum dot. Coupling of QCAD with the optimizer Dakota allows for rapid identification and improvement of device layouts that are likely to exhibit few-electron quantum dot characteristics.
\end{abstract}


%\Keywords{QCAD, Quantum dot, Poisson, Schrodinger, Predictor-corrector, Self-consistency, Optimization}

\maketitle

\section{INTRODUCTION}
Silicon-based qubits for quantum information processing have attracted enormous interest and made remarkable progress \cite{Morello2010, Borselli2011, Lim2011, Yamahata2012} during recent years due to long spin decoherence times and the well-established Si nanoelectronic fabrication infrastructure. There has been a wide range of research work \cite{Tracy2010, TMLu2011, Laird2010, Koh2012} in academia and national labs dedicated to design and fabricate Si-based multi-quantum dot (QD) devices for qubit application. However, many practical issues have made it challenging to achieve reliable few-electron multi-QD qubits, including fabrication process variation, interface quality, defect/disorder, and device design.

To facilitate the experimental development of the technology, a simulation tool was needed to efficiently solve for semiclassical (\emph{i.e.},~using the Thomas-Fermi approximation) and self-consistent quantum electrostatic potentials, single- and multi-electron wave functions, and energies at near-zero temperatures. It was also essential that such a tool be able to simulate and optimize many different single- and multi-QD structures very efficiently, and provide fast feedback on which device layouts are more likely to lead to few-electron behavior. Although there have been numerous papers \cite{Stopa1996, Milicic2000, Friesen2003, Melnikov2005, Klimeck2008} on simulation methods for quantum dots, they focus on different issues than those addressed in this work. Existing commercial \cite{SDevice2010} and academic \cite{Nanohub} device simulators either target room-temperature and many-electron devices, whereas qubit applications require temperatures close to zero Kelvin and one/few-electron devices, or target simple and few geometries, whereas multi-QD devices have very complex three-dimensional (3D) shapes and can have many different layouts due to their inherently large design space. These existing tools often have license and platform restrictions which place non-fundamental but very real limits on simulation efficiency. The Quantum Computer Aided Design (QCAD) simulator has been developed to address these challenges associated with modeling realistic multi-QDs, including complex geometries, many device layouts, low temperature operation, and 3D quantum confinement effects. It is an integrated open-source finite-element-based tool that leverages a number of Sandia-developed software programs \cite{Trilinos}, including the Trilinos suite, the Albany code \cite{Albany}, the Dakota toolbox, and the Cubit geometry and meshing tool.

The first and major part of this work details the QCAD simulator \cite{Gao2012} (or just ``QCAD'' for brevity), which is built upon the Albany finite element framework \cite{Albany} and contains three core modules of Poisson (P), Schrodinger (S), and Configuration Interaction (CI) solvers. These physical solvers can be run individually or combined self-consistently (i.e., self-consistent S-P and S-P-CI solvers) for simulating arbitrary 1D/2D/3D quantum devices made from multiple different materials. They have demonstrated fast and robust convergence behavior even at very low (milli-Kelvin) temperatures. Furthermore, very high simulation throughput has been achieved using a combination of pre- and post-processing scripting, automated structure creation and meshing, and distributed parallel computing capability and resources. In Sec.~\ref{sec:QCADSimulator}, details of the QCAD simulator and its core physical solvers are presented, with the semiclassical Poisson solver detailed in Sec.~\ref{sec:PoissonSolver}, the Schrodinger solver described in Sec.~\ref{sec:SchrodingerSolver}, and the self-consistent S-P solver detailed in Sec.~\ref{sec:SelfconSP}.

The second part of this work presents two examples of applying QCAD to simulate realistic QD structures. Since QCAD solves for the electron density at a given set of device control voltages, it is able to compute the capacitance between a quantum dot and a control gate. Such capacitances can also be determined experimentally. In Sec.~\ref{sec:DQDCap}, we compute the dot-to-gate capacitances in an experimental QD structure, using the semiclassical and self-consistent S-P solvers, and compare them with experiment. A detailed analytic analysis is then conducted to help understand the effect of quantum confinement on capacitances.  Finally, in Sec.~\ref{sec:DQDOpt}, QCAD's ability to assess and improve upon the design of nano-scale devices is demonstrated with few-electron double QDs (DQDs).  Through its coupling with the optimization driver Dakota, QCAD is able to optimize gate voltages to simultaneously achieve multiple desired targets (\emph{e.g.} few electrons in a dot and a tunnel barrier being controllable) in many different DQD device designs. We show how this aids in the identification of device designs which are likely to exhibit few-electron quantum dot behavior.

\section{QCAD SIMULATOR}
\label{sec:QCADSimulator}

\begin{figure}
\centering\includegraphics[width=8.0cm, bb=0 0 445 275]{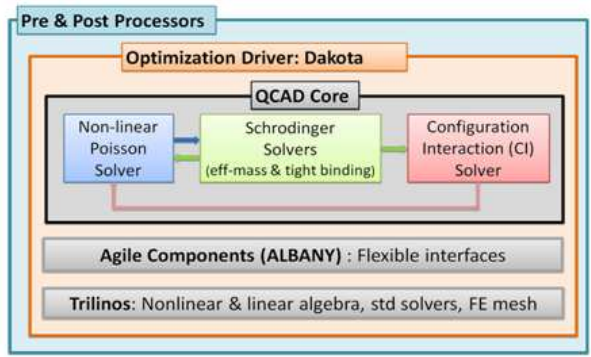}
\caption{ \label{fig:QCADiagram} (Color online) Schematic diagram of the QCAD code structure.}
\end{figure}

The QCAD code structure is given in Fig.~\ref{fig:QCADiagram}. The base of the structure is Trilinos \cite{Trilinos}, an open-source suite consisting of mathematical libraries (nonlinear and linear solvers, preconditioners, eigensolvers, etc.), discretization (finite element and finite volume) utilities, automatic differentiation library, distributed parallel library, and many more packages (refer to trilinos.sandia.gov for details). The Albany code \cite{Albany} provide a unified and flexible interface to those Trilinos packages to minimize the coding effort for users developing physical models. The Dakota driver \cite{Trilinos} interfaces with QCAD and repeatedly calls the QCAD executable to perform specified optimization tasks. It is an open-source tool that provides a broad spectrum of capabilities including nonlinear least square optimization (see dakota.sandia.gov for details). Pre- and post-processing scripts have been developed to automatically generate QCAD input files for multiple DQD devices, submit jobs to computing clusters, and collect simulation results. These scripts are one of the enabling factors that make efficient QCAD simulations possible.

The QCAD core contains nonlinear Poisson (P), effective-mass Schrodinger (S), and CI solvers. These solvers can be run individually or combined in a self-consistent manner. The Poisson and Schrodinger solvers can be coupled self-consistently to obtain single-particle envelope wave functions and energy levels. The S-P solutions can then be used as wave function basis for the CI solver which is then coupled with the Poisson solver. This self-consistent S-P-CI technique produces accurate many-particle wave functions and energies within the effective mass approximation, which are important for quantitative study of few-electron multi-QD behavior (\emph{e.g.}~for computing the exchange energy in DQDs). Details about the Poisson, Schrodinger, and self-consistent S-P solvers are described in the following subsections, and the CI method can be found in Ref.~\onlinecite{NielsenCI2010}.  We note that the current version of the QCAD simulator solves one or a system of equations in the full number of dimensions of the problem, and does not make any separability assumptions that allow a higher dimensional problem (e.g.~3D) to be divided into multiple lower dimensional problems (e.g.~2D+1D). Such simplifying approximations\cite{Anderson_2009} are a possible future enhancement to the QCAD core.

\subsection{Poisson Solver}
\label{sec:PoissonSolver}

The well-known Poisson equation in a bulk semiconductor is given by
\begin{equation}
\label{eq:Poisson}
-\nabla \cdot (\epsilon_{s} \nabla \phi) = q (p - n + N_{D}^{+} - N_{A}^{-}),
\end{equation}
where $\phi$ is the electrostatic potential, $\epsilon_{s}$ is the static permittivity, $q$ is the elementary charge, $n$ and $p$ are the electron and hole concentrations respectively, $N_{D}^{+}$ and $N_{A}^{-}$ are ionized donor and acceptor concentrations respectively. Note that the form of the left hand side in Eq.~(\ref{eq:Poisson}) allows $\epsilon_{s}$ to have spatial dependence.

\subsubsection{Carrier Statistics}

$n$ and $p$ are given by carrier statistics for bulk (spatially unconfined) semiconductors. Both Maxwell-Boltzmann (MB) and Fermi-Dirac (FD) statistics are implemented in QCAD. For the MB statistics, $n$ and $p$ take the exponential forms,
\begin{eqnarray}
\label{eq:MBstatistics}
n &=& N_{C} \exp \biggl( \frac{E_{F}-E_{C}}{k_{B}T} \biggl), \nonumber\\
p &=& N_{V} \exp \biggl( \frac{E_{V}-E_{F}}{k_{B}T} \biggl),
\end{eqnarray}
where $k_{B}$ is the Boltzmann constant, $T$ is the lattice temperature, $E_C$ and $E_V$ are the conduction and valence band edge respectively, and $E_F$ is the extrinsic Fermi level (more details on $E_C$, $E_V$, and $E_F$ are given in Sec.~\ref{sec:refPotential}). For the FD statistics, $n$ and $p$ are expressed in terms of the Fermi-Dirac integrals (see Appendix A for the derivation),
\begin{eqnarray}
\label{eq:FDstatistics}
n &=& N_{C} \mathcal{F}_{1/2} \biggl( \frac{E_{F}-E_{C}}{k_{B}T} \biggl), \nonumber\\
p &=& N_{V} \mathcal{F}_{1/2} \biggl( \frac{E_{V}-E_{F}}{k_{B}T} \biggl).
\end{eqnarray}
$N_{C}$ and $N_{V}$ are effective density of states (DOS) in the conduction and valence band, respectively. Assuming parabolic band structure, we have
\begin{eqnarray}
\label{eq:EffectiveDOS}
N_{C} &=& 2 \biggl( \frac{m^{*}_{n} k_{B} T}{2 \pi \hbar^2} \biggl)^{3/2}, \nonumber\\
N_{V} &=& 2 \biggl( \frac{m^{*}_{p} k_{B} T}{2 \pi \hbar^2} \biggl)^{3/2},
\end{eqnarray}
where $\hbar$ is the reduced Planck constant, $m^{*}_{n}$ and $m^{*}_{p}$ are respectively the electron and hole DOS effective mass including all equivalent band minima. For bulk silicon, there are six equivalent conduction minima, and the valence band minimum is degenerate including heavy hole and light hole bands at the $\Gamma$ valley, hence
\begin{eqnarray}
\label{eq:EffectiveMass}
m^{*}_{n} &=& 6^{2/3} (m_{l} m_{t}^2)^{1/3}, \nonumber\\
m^{*}_{p} &=& \biggl( m_{hh}^{3/2} + m_{lh}^{3/2} \biggl)^{2/3},
\end{eqnarray}
with $m_{l}$, $m_{t}$, $m_{hh}$, and $m_{lh}$ being the electron longitudinal, electron transverse, heavy hole, and light hole effective mass, respectively.

\begin{figure}
\centering\includegraphics[width=8.0cm, bb=80 215 515 560]{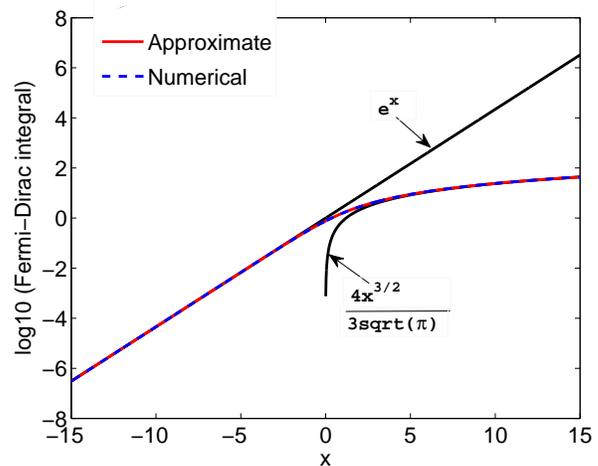}
\caption{ \label{fig:FDIntCmprsn} (Color online) Fermi-Dirac integral of 1/2 order in
logarithmic scale. Black curves are the asymptotic expansions, $\exp(x)$ for $x \rightarrow -\infty$, and $\frac{4x^{3/2}} {3 \sqrt{\pi}}$ for $x \rightarrow +\infty$. The red solid curve labeled as Approximate is obtained using Eq.~(\ref{eq:FDIntApprox}), while the blue dashed curve labeled as Numerical uses the numerical integration method in Ref.~\onlinecite{Mohan1995}. The red solid and blue dashed curves do not show any visible difference, and agree well with the asymptotic forms for large $|x|$, whereas they differ significantly from the asymptotic forms in the region of $x \in (-4, 4)$. }
\end{figure}

The $\mathcal{F}_{1/2}(x)$ function in Eq.~({\ref{eq:FDstatistics}) is the Fermi-Dirac integral of $1/2$ order and is defined as \cite{Blakemore1982}
\begin{equation}
\label{eq:FDIntegral}
\mathcal{F}_\frac{1}{2}(x) = \frac{2}{\sqrt{\pi}} \int_{0}^{\infty} \frac{\sqrt {\varepsilon} \mathrm{d} \varepsilon}
{1+\exp ( \varepsilon - x )}.
\end{equation}
Although the closed form of this integral can be formally expressed by the polylogarithm function\cite{PolyLogarithm} or by a complete expansion discussed in Ref.~\onlinecite{Kim2008}, the polylogarithm function and the complete expansion involve summations of infinite series.  Hence in practice, one has to either use certain approximation to obtain a computable analytic expression, or use numerical integration techniques\cite{Mohan1995, Press2007}. There have been a few approximate analytic forms proposed in literature ~\cite{Bedn1978, Blakemore1982, Halen1985} that offer relatively simple expressions and sufficient accuracy. Among them, the approximate expression in Ref.~\onlinecite{Bedn1978} takes a single simple form and provides a relative error less than $0.4\%$ for $x \in ({-\infty, +\infty})$, hence has been widely used in the device modeling community \cite{VasileskaNanohub}. The expression in Ref.~\onlinecite{Bedn1978} is given as
\begin{eqnarray}
\label{eq:FDIntApprox}
\mathcal{F}_\frac{1}{2}(x) &\approx& \biggl( e^{-x} + \frac{3 \sqrt{\pi}}{4} v^{-3/8} \biggl)^{-1}, \nonumber \\
v &=& x^4 + 50 + 33.6 x (1-0.68 \exp [-0.17(x+1)^2]. \nonumber \\
\end{eqnarray}
The asymptotic expansion at $x \rightarrow -\infty$ leads to $\mathcal{F}_{1/2}(x) = \exp(x)$, implying that Eq.~(\ref{eq:FDstatistics}) becomes equivalent to Eq.~(\ref{eq:MBstatistics}), which is the case for non-degenerate semiconductors where $E_F \ll (E_C -$ [a few $k_{B}T$]) for $n$, and $E_F \gg (E_V + $ [a few $k_{B}T$]) for $p$. The asymptotic form at $x \rightarrow +\infty$ is $\mathcal{F}_{1/2}(x) = \frac{4x^{3/2}} {3 \sqrt{\pi}}$, which corresponds to the strongly degenerate case near 0 K, where $E_F \gg (E_C + $ [a few $k_{B}T$]) for $n$ (i.e., the Fermi level is located within the conduction band), and $E_F \ll (E_V - $[a few $k_{B}T$]) for $p$ (i.e., the Fermi level is inside the valence band). Figure~\ref{fig:FDIntCmprsn} shows a comparison of the 1/2-order Fermi-Dirac integral calculated by different methods. It is clear that the approximate expression in Eq.~(\ref{eq:FDIntApprox}) produces visually the same result as the numerical approach \cite{Mohan1995}, and follows the proper asymptotic forms for large $|x|$. In the small $|x|$ regime, neither of the asymptotic forms is valid. Since semiconductor DQD qubits are currently operated in this regime (corresponding to very low temperatures, mK to a couple of K), it is important to adopt a sufficiently accurate evaluation of the Fermi-Dirac integral. Due to the good accuracy and simplicity of Eq.~(\ref{eq:FDIntApprox}), we implemented this form in QCAD for the FD statistics. In the actual implementation, we approximate $ \mathcal{F}_{1/2}(x)$ by $ e^x $ for $x < -50$ to avoid numerical instability caused by the $e^{-x}$ term in Eq.~(\ref{eq:FDIntApprox}). Such large negative values of $x$ can occur at very low temperatures, and this approximation results in no discernible loss of accuracy, as shown in Fig.~\ref{fig:FDIntCmprsn}.

\subsubsection{The Reference Potential}
\label{sec:refPotential}

Before solving the Poisson Eq.~(\ref{eq:Poisson}) for the electrostatic potential $\phi$, one needs to relate $\phi$ to the band energies of the materials making up the device. One requirement for the electrostatic potential is that it must be continuous everywhere in a device. For a homo-junction device such as a PN silicon diode, $E_C$, $E_V$, and $E_i$ (the intrinsic Fermi level) as functions of position are parallel to each other and continuous across the device, so it is natural to choose $-q\phi = E_{i}$, i.e., to solve for the inverse of intrinsic Fermi level. In an arbitrary hetero-junction structure, however, $E_C$, $E_V$, and $E_i$ could all be discontinuous. Figure \ref{fig:qPhiRefDefinition} shows a schematic of the band structure of a MOS-type device under zero bias illustrating the discontinuity of $E_C$ and $E_V$. What \emph{is} always continuous in arbitrary homo- and hetero-junction devices is the vacuum level indicated as $E_0$ in Fig.~\ref{fig:qPhiRefDefinition}. Therefore, we choose $\phi$ to satisfy \cite{Lundstrom1981}
\begin{equation}
\label{eq:qPhiRef}
-q (\phi - \phi_{ref}) = E_0 = E_C + \chi,
\end{equation}
where $\phi_{ref}$ is a constant reference potential and $\chi$ is the electron affinity of a material. This choice implies that we are solving for the inverse of the vacuum level shifted by a constant value. While in theory different $\phi_{ref}$ values only change the resulting solution, $\phi$, by a constant offset, in practice they can lead to different numerical convergence behavior during simulation. A good choice of $\phi_{ref}$ that has shown numerical robustness in devices containing silicon is to select as the reference potential the intrinsic Fermi level of silicon relative to the vacuum level, i.e., $q\phi_{ref} = E_0 - E_i$(Si).  For a more detailed explanation of band diagrams schematics such as that in Fig.~\ref{fig:qPhiRefDefinition}, we refer the reader to Ref.~\onlinecite{Tan_SchroPo_1990}.

\begin{figure}
\centering\includegraphics[width=6.0cm, bb=185 320 400 505]{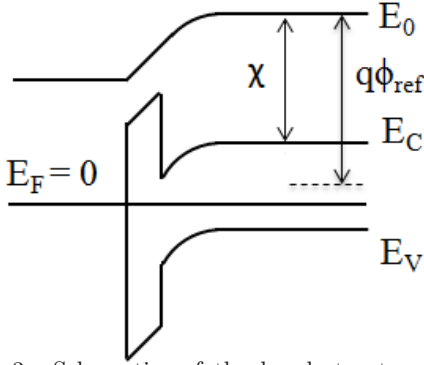}
\caption{ \label{fig:qPhiRefDefinition} Schematics of the band structure of a MOS-type device under zero bias. }
\end{figure}

With the choice of $\phi$ in Eq.~(\ref{eq:qPhiRef}), we can rewrite $E_C$ and $E_V$ as
\begin{eqnarray}
\label{eq:EcEv_phi}
E_C &=& -q (\phi - \phi_{ref}) - \chi, \nonumber\\
E_V &=& -q (\phi - \phi_{ref}) - \chi - E_g,
\end{eqnarray}
with $E_g$ being the band gap of a material. Then the $n$ and $p$ in Eq.~(\ref{eq:Poisson}) becomes a function of $\phi$ only, i.e., $n(\phi)$ and $p(\phi)$, as $\chi$ and $E_g$ are material-dependent parameters and known for most semiconductors.

As QCAD does not solve the carrier transport (e.g., drift-diffusion) equations, all calculations must assume that thermal equilibrium (zero current flow) has been attained. The Fermi level, $E_F$, is taken to be a constant throughout any electrically-connected region of a device.  The value of this constant is set by the voltages applied to the device.  For example, if a voltage $V_{sub}$ is applied to the substrate (right side) of the device in Fig.~\ref{fig:qPhiRefDefinition}, $E_F$ will become $-qV_{sub}$. The band structure shown in Fig.~\ref{fig:qPhiRefDefinition} corresponds to a MOS-type structure with metal gate (left side). If the structure instead had a highly-doped semiconductor gate (e.g., n$^{+}$ polysilicon gate) with applied voltage $V_g$, $E_F$ in the gate region would be $-qV_g$ while $E_F$ in the substrate region remains at $-qV_{sub}$.

\subsubsection{Incomplete Ionization}

When impurities are introduced into the semiconductor crystals, depending on the impurity energy level and the lattice temperature, not all dopants are necessarily ionized, especially at very low lattice temperatures (where DQD qubits are commonly operated).  The ionized concentration for donors and acceptors is given by \cite{SzeBook}
\begin{eqnarray}
\label{eq:Ionization}
N_{D}^{+} &=& \frac {N_D} {1 + g_D \exp(\frac{E_F - E_D}{k_B T})}, \nonumber\\
N_{A}^{-} &=& \frac {N_A} {1 + g_A \exp(\frac{E_A - E_F}{k_B T})},
\end{eqnarray}
where $E_D$ is the donor energy level, $g_D$ is the donor ground state degeneracy factor, $E_A$ is the acceptor energy level, and $g_A$ is the acceptor ground state degeneracy factor. $g_D$ is equal to 2 because a donor level can accept one electron with either spin or can have no electron when filled. $g_A$ is equal to 4 because in most semiconductors each acceptor level can accept one hole of either spin and the impurity level is doubly degenerate as a result of the two degenerate valence bands (heavy hole and light hole bands) at the $\Gamma$ point.

To write $N_D^{+}$ and $N_A^{-}$ as a function of $\phi$, we need to rewrite $E_F - E_D = E_F - E_C + E_C - E_D = E_F - E_C + E_d$ with $E_d$ being the donor ionization energy, and $E_A - E_F = E_A - E_V + E_V - E_F = E_a + E_V - E_F$ with $E_a$ being the acceptor ionization energy. The most common donors in bulk Si are phosphorus (P) and arsenic (As), which have ionization energies of $E_d = 46 $ meV and 54 meV respectively\cite{SzeBook}.  The most common acceptor dopant in bulk Si is boron (B), which has $E_a = 44$ meV\cite{SzeBook}.

Substituting Eq.~(\ref{eq:EcEv_phi}) and $E_F - E_D$, $E_A - E_F$ into Eq.~(\ref{eq:Ionization}), we get \cite{Trellakis2000}
\begin{eqnarray}
\label{eq:FinalIonization}
N_{D}^{+} &=& \frac {N_D} {1 + g_D \exp \biggl( \frac{E_F + E_d - q \phi_{ref} + \chi + q \phi} {k_B T} \biggl)}, \nonumber\\
N_{A}^{-} &=& \frac {N_A} {1 + g_A \exp \biggl( \frac{- E_F + E_a + q \phi_{ref} - \chi - E_g - q \phi} {k_B T} \biggl)}.
\end{eqnarray}
With these expressions, $N_{D}^{+}$ and $N_{A}^{-}$ also become a function of $\phi$, i.e., $N_{D}^{+}(\phi)$ and $N_{A}^{-}(\phi)$. Hence, the entire right hand side (RHS) of Eq.~(\ref{eq:Poisson}) can be written as a nonlinear function of $\phi$. Applying integration by parts and divergence theorem, we then rewrite the equation into the finite element (FE) weak form,
\begin{eqnarray}
\label{eq:PoissonWeakForm}
&& \int \epsilon_{s} \nabla \phi \cdot \nabla w d \Omega - \int_{\Gamma} \epsilon_{s} \nabla \phi \cdot \hat{\eta} w d \Gamma \nonumber \\
&& - \int q [ p(\phi) - n(\phi) + N_D^{+}(\phi) - N_A^{+}(\phi) ] w d \Omega = 0,
\end{eqnarray}
where $w$ is the FE nodal basis function and the second term is a line integral over the simulation domain boundary with $\hat{\eta}$ being the unit normal vector of the surface element $d\Gamma$. The weak form is discretized using the Trilinos/Intrepid library, and the resulting discrete equation is solved by a nonlinear Newton solver also in Trilinos. Both the discretization library and the Newton solver were made directly available to QCAD through the Albany framework (cf.~Fig.~\ref{fig:QCADiagram}).

\subsubsection{Boundary Conditions}

An essential ingredient to the formulation of a differential equation are boundary conditions (BCs). QCAD supports three types of BCs: Dirichlet, Neumann, and Robin BCs.  We will next discuss the implementation of these types in turn.

Dirichlet BCs are divided into two cases: (1) setting a voltage on the surface of a metallic region that borders insulator, and (2) setting a voltage on the surface of an Ohmic contact region which borders semiconductor. Case (1) is used for gate electrodes in field effect transistor (FET)-like devices, and the Dirichlet BC value $\phi_{ins}$ on the bordering insulator(s) is given by the simple expression
\begin{equation}
\label{eq:ContOnIns}
\phi_{ins} = V_{g} - \frac { \Phi_{m} - q \phi_{ref}} {q},
\end{equation}
with $V_{g}$ being the applied gate voltage and $\Phi_{m}$ being the metal work function.

In the second case (used for Ohmic contacts in semiconductors), the potential on the bordering semiconductor surfaces is computed assuming thermal equilibrium and charge neutrality at the contacts. The calculation depends on carrier statistics and dopant ionization. For MB statistics, the charge neutrality $n + N_A^{-} = p + N_D^{+}$ condition leads to
\begin{eqnarray}
\label{eq:BC4MBqNeutral}
&& N_C \exp \biggl( \frac{E_F + q \phi - q \phi_{ref} + \chi} {k_B T} \biggl) + N_A^{-}  \nonumber \\
&& = N_V \exp \biggl( \frac{ -E_F -q \phi + q \phi_{ref} - \chi - E_g} {k_B T} \biggl)+ N_D^{+}.
\end{eqnarray}
With complete ionization of dopants (i.e., $N_A^{-} = N_A$ and $N_D^{+} = N_D$ ), the potentials at n-type and p-type Ohmic contacts are respectively given by
\begin{eqnarray}
\label{eq:PotBC4MBCompIon}
\phi_{ohm}^{n} &=& \frac {q \phi_{ref} - \chi} {q} + \frac{k_B T}{q} \ln \biggl( \frac{N_D}{N_C} \biggl) + V_a, \nonumber\\
\phi_{ohm}^{p} &=& \frac {q \phi_{ref} - \chi - E_g} {q} - \frac{k_B T}{q} \ln \biggl( \frac{N_A}{N_V} \biggl) + V_a,
\end{eqnarray}
where $V_a$ is an externally applied voltage. When including incomplete ionization effect of dopants, we have for n-type and p-type semiconductors respectively,
\begin{eqnarray}
\label{eq:BC4MBIncompIon}
N_C \exp \biggl( \frac{E_F - E_C} {k_B T} \biggl) = \frac{N_D} {1 + 2 \exp \biggl( \frac{E_F - E_D} {k_B T} \biggl) } , \nonumber \\
N_V \exp \biggl( \frac{E_V - E_F} {k_B T} \biggl) = \frac{N_A} {1 + 4 \exp \biggl( \frac{E_A - E_F} {k_B T} \biggl) } .
\end{eqnarray}
Let us denote $y_n = \exp ( \frac{E_F - E_D} {k_B T} )$ and $y_p = \exp ( \frac{E_A - E_F} {k_B T} )$. Then, by the definitions of $E_a$ and $E_d$, we obtain the identities, $\exp ( \frac{E_F - E_C} {k_B T} ) = y_n \exp ( \frac{-E_d} {k_B T} )$ and $\exp ( \frac{E_V - E_F} {k_B T} ) = y_p \exp ( \frac{-E_a} {k_B T} ) $. Substituting the identities into Eq.~(\ref{eq:BC4MBIncompIon}), we obtain
\begin{eqnarray}
\label{eq:Yval4MBIncompIon}
y_n &=& -\frac{1}{4} + \frac{1}{4} \biggl[ 1 + \frac{8 N_D}{N_C} \exp \biggl( \frac{E_d} {k_B T} \biggl) \biggl] ^{1/2}, \nonumber \\
y_p &=& -\frac{1}{8} + \frac{1}{8} \biggl[ 1 + \frac{16 N_A}{N_V} \exp \biggl( \frac{E_a} {k_B T} \biggl) \biggl] ^{1/2}.
\end{eqnarray}
From the definitions of $y_n$ and $y_p$, the use of $-q V_a = E_F$, and Eq.~(\ref{eq:EcEv_phi}), we can obtain the potentials that include dopant incomplete ionization effect at the n-type and p-type Ohmic contacts respectively as, \begin{eqnarray}
\label{eq:PotBC4MBIncompIon}
\phi_{ohm}^{n} &=& \frac{-E_d + q \phi_{ref} - \chi} {q} + \frac{k_B T}{q} \ln (y_n) + V_a , \nonumber \\
\phi_{ohm}^{p} &=& \frac{-E_a + q \phi_{ref} - \chi - E_g} {q} - \frac{k_B T}{q} \ln (y_p) + V_a. \nonumber \\
\end{eqnarray}
At very low temperatures, the exponential terms in Eq.~(\ref{eq:Yval4MBIncompIon}) could blow up numerically. To avoid numerical instability in QCAD, we approximate the $\ln (y_n)$ and $\ln (y_p)$ terms for very low temperatures as,
\begin{eqnarray}
\label{eq:ApproxLogTerms}
\ln (y_n) &=& \frac{1}{2} \ln \biggl( \frac{N_D} {2 N_C} \biggl) + \frac{E_d}{2 k_B T}, \nonumber \\
\ln (y_p) &=& \frac{1}{2} \ln \biggl( \frac{N_A} {4 N_V} \biggl) + \frac{E_a}{2 k_B T}.
\end{eqnarray}

For FD statistics and assuming complete ionization of dopants, we have
\begin{eqnarray}
\label{eq:BC4FDCompIon}
N_C \mathcal{F}_\frac{1}{2} \biggl( \frac{E_F - E_C}{k_B T}  \biggl) &=& N_D \; \; \; \text{for n-type},  \nonumber \\
N_V \mathcal{F}_\frac{1}{2} \biggl( \frac{E_V - E_F}{k_B T}  \biggl) &=& N_A \; \; \; \text{for p-type}.
\end{eqnarray}
To solve for the potentials at Ohmic contacts, in principle, we need to numerically solve Eq.~(\ref{eq:BC4FDCompIon}) as the Fermi-Dirac integral does not have an analytic result. In QCAD, we use an approximate expression for the inverse of the 1/2 order Fermi-Dirac integral, that is, given $u = \mathcal{F}_{1/2}(\eta)$, $\eta$ is computed as \cite{Nilsson1973},
\begin{eqnarray}
\label{eq:FDintInverse}
\eta &=& \frac{- \ln (u)} {u^2 -1} + \frac{\nu} {1+ (0.24 + 1.08 \nu)^{-2}}, \nonumber \\
\nu &=& \biggl( \frac {3 \sqrt \pi u} {4} \biggl) ^{2/3}.
\end{eqnarray}
This approximation has an error of less than $0.6 \% $ for the entire $\eta$ range. Using this expression, we obtain the BC potentials as,
\begin{eqnarray}
\label{eq:PotBC4FDCompIon}
\phi_{ohm}^{n} &=& \frac{q \phi_{ref} - \chi} {q} + \frac{k_B T}{q} \eta  + V_a, \nonumber \\
\phi_{ohm}^{p} &=& \frac{q \phi_{ref} - \chi - E_g} {q} - \frac{k_B T}{q} \eta  + V_a,
\end{eqnarray}
with $\eta$ given in Eq.~(\ref{eq:FDintInverse}) where $u = N_D / N_C$ for n-type and $u = N_A / N_V$ for p-type. For the case of FD statistics and incomplete ionization, there exists no approximate analytic expressions for the BC potentials, and one has to solve a non-trivial nonlinear equation if want to be very accurate. In QCAD, we approximate this case using MD statistics with incomplete ionization and utilize the BC potentials given in Eq.~(\ref{eq:PotBC4MBIncompIon}).

Neumann BCs in finite element methods are used to specify how ``flux'' is conserved across boundaries. By default, all boundaries that are not given any other type of boundary conditions, assume implicit Neumann BCs which preserve the flux. In the case of the Poisson equation, the flux is $\epsilon_s \nabla \phi \cdot \hat{\eta}$, where $\hat{\eta}$ is the unit normal of the boundary surface.  Thus, by default (i.e.~when no other boundary condition is specified), $\epsilon_{s} \nabla \phi \cdot \hat{\eta} = 0$ on outer boundaries of the finite element mesh and $\epsilon_{s1} \nabla \phi_1 \cdot \hat{\eta}_1 = \epsilon_{s2} \nabla \phi_2 \cdot \hat{\eta}_2 $ on internal boundaries.  These two conditions are automatically satisfied in the finite element framework by setting the $\int \epsilon_{s} \nabla \phi \cdot \hat{\eta} w d \Gamma$ term to 0 in Eq.~(\ref{eq:PoissonWeakForm})

QCAD has the ability to specify non-flux-conserving Neumann BCs on specific boundaries such that the difference between the fluxes on either side of the boundary are equal to some specified constant value.  Written mathematically, $(\epsilon_{s2} \nabla \phi_2 - \epsilon_{s1} \nabla \phi_1) \cdot \hat{\eta} = q \sigma_s $ , where $\hat{\eta}$ is the unit normal vector of the interface pointing from material 2 to 1 and $\sigma_s$ is the specified constant.  Physically, $\sigma_s$ is a surface charge density located at the boundary.  Note that when $\sigma_s=0$ (i.e. no surface charge), the boundary condition reduces to the default flux-conserving condition.  Within the finite element discretization in QCAD, this type of Neumann BC is implemented in the integral form
\begin{equation}
\label{eq:NeuBC4fc}
\int_{\Gamma_{cbc}} (\epsilon_{s2} \nabla \phi_2 - \epsilon_{s1} \nabla \phi_1) \cdot \hat{\eta} w d \Gamma_{cbc}
= \int_{\Gamma_{cbc}} q \sigma_s w d \Gamma_{cbc}.
\end{equation}

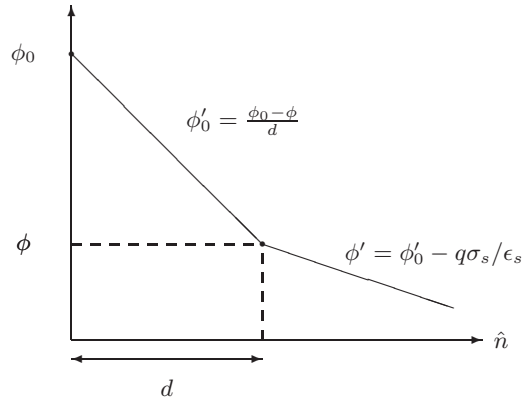
\begin{figure}
\setlength{\unitlength}{1in}
\begin{center}
\begin{picture}(3,2.5)
\put(0.5,0.5){\vector(1,0){2.15}}
\put(0.5,0.5){\vector(0,1){1.75}}
\put(0.5,2){\circle*{0.03}}
\put(0.5,2){\line(1,-1){1}}
\put(1.5,1){\circle*{0.03}}
\put(1.5,1){\line(3,-1){1}}
\multiput(0.5,1)(0.1,0){10}{\line(1,0){0.05}}
\multiput(1.5,0.5)(0,0.1){5}{\line(0,1){0.05}}

\put(1,0.4){\vector(1,0){.5}}
\put(1,0.4){\vector(-1,0){.5}}
\put(1,0.25){\makebox(0, 0){$d$}}

\put(2.75,0.5){\makebox(0, 0){$\hat{n}$}}
\put(0.25,2){\makebox(0, 0){$\phi_0$}}
\put(0.25,1){\makebox(0, 0){$\phi$}}
\put(0.25,1){\makebox(0, 0){$\phi$}}
\put(1.4,1.65){\makebox(0, 0){$\phi'_0 = \frac{\phi_0-\phi}{d}$}}
\put(2.4,0.95){\makebox(0, 0){$\phi' = \phi'_0 - q\sigma_s/\epsilon_s$}}
\end{picture}
\caption{~The derivation of the Robin BC parameters used to set a value of the potential and a nearby surface charge on a single surface. This diagram shows a 1D cut along the direction $\hat{n}$ normal to a surface element. $\phi_0$, $\sigma_s$, and $d$ are given values, and $\phi$ is the variable being solved.  The derivative $\phi'$ is along the direction normal to the surface.  Combining the equations yields $\epsilon_s \phi' = \epsilon_s \frac{\phi_0-\phi}{d} - q\sigma_s$, which takes the form of a Robin boundary condition.\label{fig:RobinBCDiagram}}
\end{center}
\end{figure}

A major shortcoming of the Neumann BCs is their inability to characterize surface charge on (or extremely close to) an interface whose voltage is set by a Dirichlet BC.  This is due to the simple fact that specifying both Dirichlet and Neumann BCs on the same surface over-determines the problem.  Yet, this is essentially what is needed to model a layer of charge that is stuck to one of the conducting gates (often polysilicon) used to control a device.  One way around this technical difficulty is to place a layer of very thin finite-element cells around the charged gate and set a Neumann BC on the new surface lying a small distance away from the gate itself.  This approach, however, suffers due to the thin finite elements adversely affecting convergence and their being hard to create in the first place.  Instead, we use what are called Robin boundary conditions\cite{RobinBCs} to address the issue of charged gates. Robin BCs are similar to Neumann BCs but allow the flux at a surface to depend on the value of the solution (in this case the potential) there. Specifically, the Robin BC for an internal surface element can be written $(\epsilon_{s2}\nabla\phi_2 - \epsilon_{s1}\nabla\phi_1)\cdot\hat{\eta} = C + \alpha(\phi_2-\phi_1)$, where $C$ and $\alpha$ are fixed constants.  At an external surface, we have $\epsilon_s\nabla\phi\cdot\hat{\eta} = C + \alpha\phi$.  We would like to roll into a single boundary condition, a Dirichlet condition at one surface followed by a Neumann boundary condition at a parallel surface lying a very small distance away from the first surface.  This can be done at an external surface using $C=\epsilon_s\phi_0/d-q\sigma_s$ and $\alpha=\epsilon_s/d$, as shown in Fig.~\ref{fig:RobinBCDiagram}, which places a surface charge of $q\sigma_s$ a distance $d$ away from a point at which $\phi$ is pinned to $\phi_0$.  We somewhat arbitrarily choose $d=10\,\mbox{nm}$, which is much smaller than any of the mesh features (for semiclassical Poisson simulations) in our devices of interest. Robin BCs are enforced in an integral form, similar to that of Neumann BCs (cf.~Eq.~\ref{eq:NeuBC4fc}).

\subsection{Schrodinger Solver}
\label{sec:SchrodingerSolver}

\begin{figure}
\centering
\includegraphics[width=0.85\linewidth, bb=80 215 520 560]{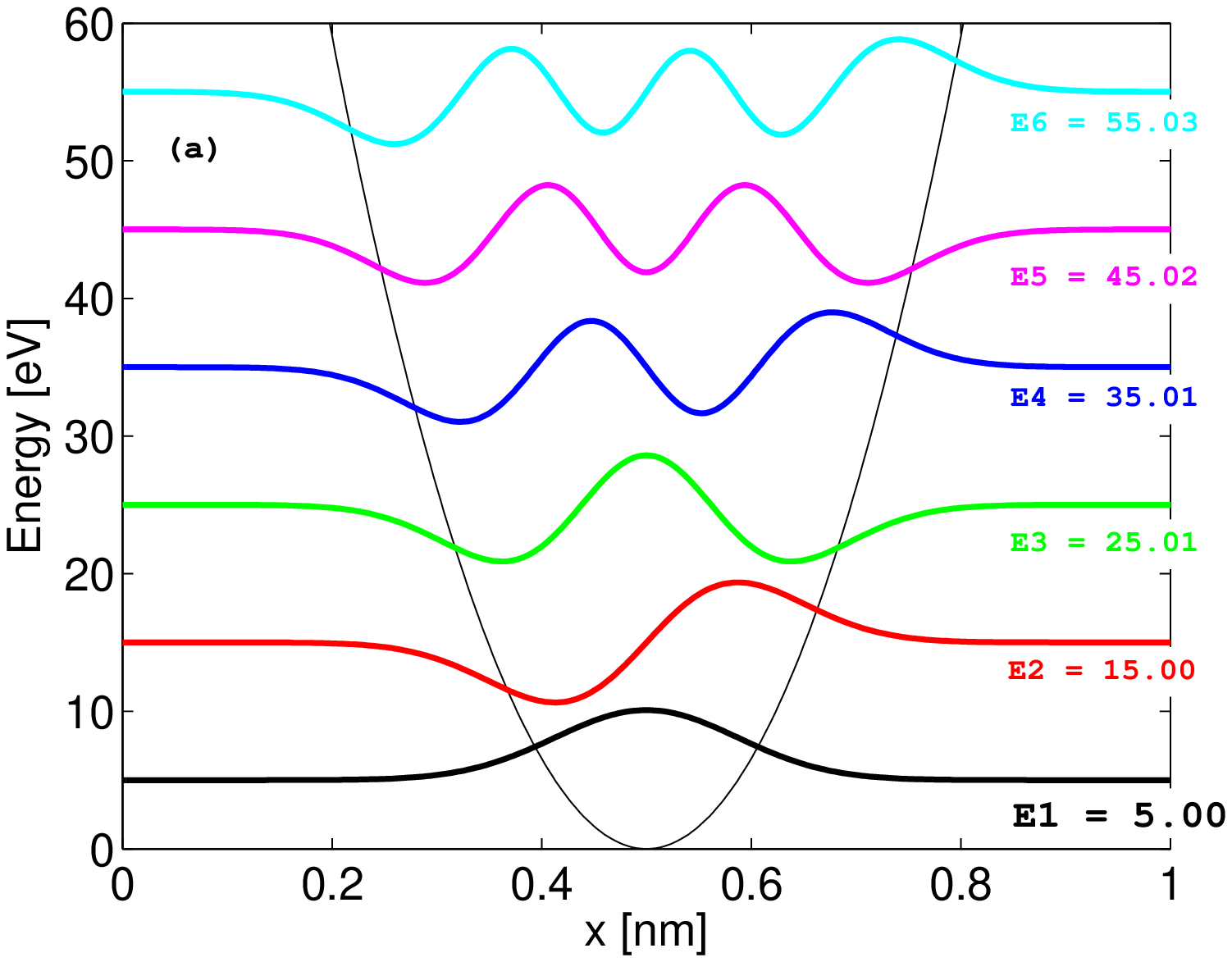}
\includegraphics[width=0.85\linewidth, bb=80 215 520 565]{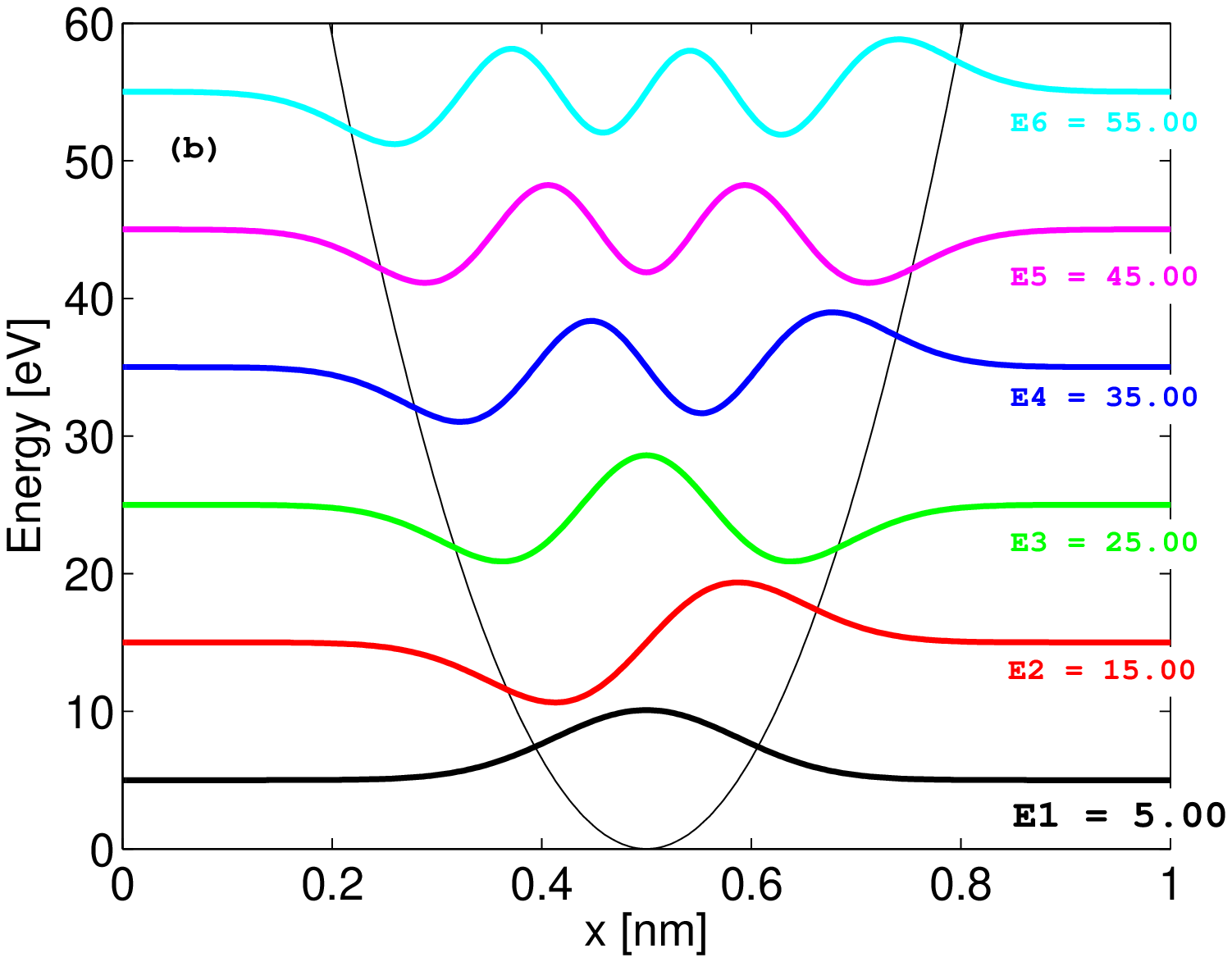}
\caption{ \label{fig:ParabolaWFs} (Color online) (a) Wave functions and energies obtained from the QCAD Schrodinger solver for a 1D parabolic potential well. (b) Analytic wave functions and energies for the same potential well. All the wave functions are scaled by the same factor for easy visualization. It is clear that QCAD wave functions and energies agree very well with the analytic results. }
\end{figure}

The time-independent single-particle effective mass Schrodinger equation takes the form of
\begin{equation}
\label{eq:SchrodEqn}
\frac{-\hbar^2}{2} \nabla \biggl( \frac{1}{m^*} \nabla \psi(\textbf{r}) \biggl) + V(\textbf{r}) \psi(\textbf{r}) = E \psi(\textbf{r}).
\end{equation}
The FE weak form of the equation is
\begin{eqnarray}
\label{eq:SchrodEqnWeakForm}
&& \frac{\hbar^2}{2 m^*} \biggl( \int \nabla \psi \cdot \nabla w d \Omega - \int_{\Gamma} \nabla \psi \cdot \hat{\eta} w d \Gamma \biggl) \nonumber \\
&& + \int V \psi w d \Omega - \int E \psi w d \Omega = 0
\end{eqnarray}
The weak form is discretized by the FE method and the resulting eigenvalue problem $[H] [\psi] = [E] [\psi]$ is solved by the Trilinos eigensolver package called Anasazi \cite{Trilinos}.

The Schrodinger solver supports two types of boundary conditions: Dirichlet and Neumann. For Dirichlet boundaries, $\psi = 0$. All other boundaries excluding Dirichlet are treated as Neumann BCs which, require $\frac{1}{m^*} \nabla \psi \cdot \hat{\eta} = 0$ on outer boundaries, and $\frac{1}{m^*} \nabla \psi \cdot \hat{\eta} $ being continuous (i.e., flux conservation) across material interfaces on internal boundaries. As in the Poisson solver, Neumann BCs are automatically satisfied in the FE framework by setting $\int_{\Gamma} \nabla \psi \cdot \hat{\eta} w d \Gamma = 0$ in Eq.~(\ref{eq:SchrodEqnWeakForm}). (The ability to set non-flux-conserving Neumann boundary conditions is absent in the Schrodinger solver since it would have no application in our work.)

Figure \ref{fig:ParabolaWFs} shows a comparison between the QCAD Schrodinger solver and analytic results of the lowest six wave functions and energies for a 1D parabolic potential well. The QCAD and analytic results are in excellent agreement. The solver was also applied to 2D and 3D infinite potential wells. The obtained wave functions and energies were compared with the analytic results and excellent agreement was also observed.

\begin{figure}
\centering
\includegraphics[width=0.85\linewidth, bb=75 215 515 565]{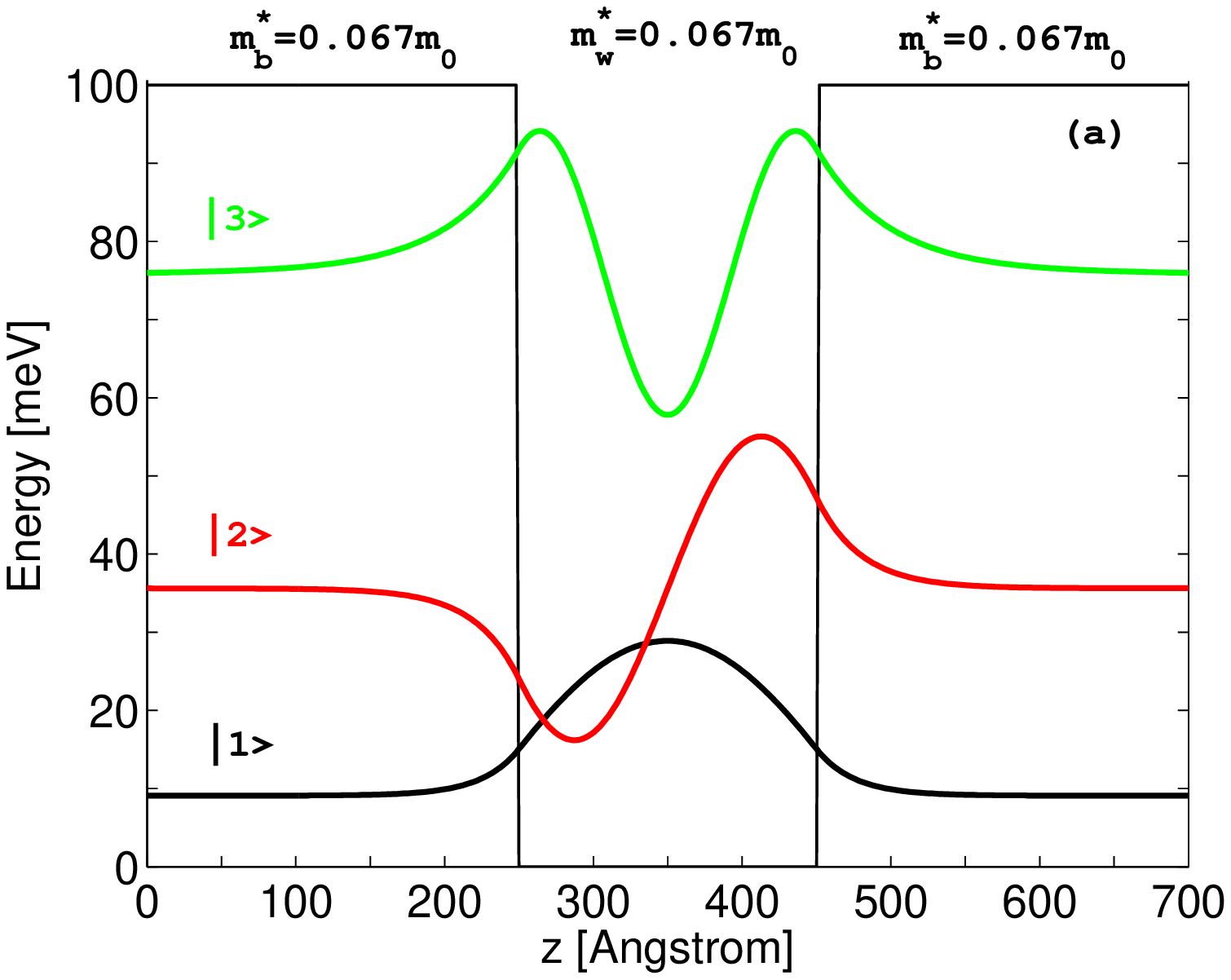}
\includegraphics[width=0.85\linewidth, bb=75 215 515 570]{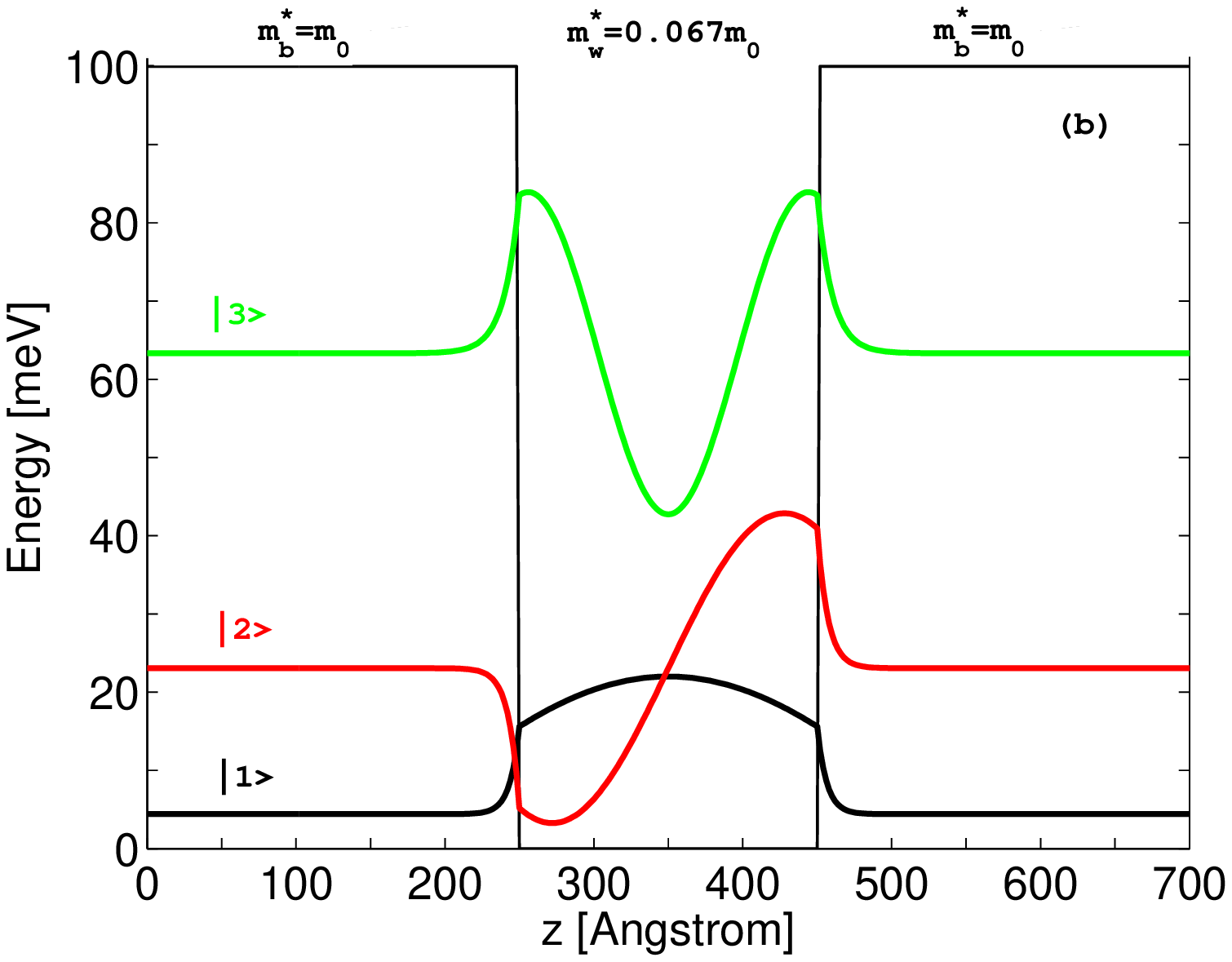}
\caption{ \label{fig:FiniteWFs} (Color online) (a) Lowest three wave functions and energies obtained from the QCAD Schrodinger solver for the 1D finite potential well with $m_b^* = m_w^* = 0.067 m_0$ where $m_0$ is the free electron mass. (b) Same as (a) except $m_b^* = m_0$ and $m_w^* = 0.067 m_0$. All wave functions are scaled by the same factor for easy visualization. }
\end{figure}

One of the advantages of using the FE discretization over the finite difference discretization is that the continuities of $\psi$ and $\frac{1}{m^*} \nabla \psi$ across heterojunctions are automatically satisfied in the former case, whereas they have to be explicitly enforced in the latter case. Specifically, when going from a homojunction device to a heterojunction device, the QCAD Schrodinger solver does not require any code change except setting the proper effective masses for the materials used. As an example, consider a 1D finite potential well that has a width of 20 nm, a potential height of 100 meV, and can have different effective masses for the well and barrier. Figure \ref{fig:FiniteWFs}(a) shows the lowest three wave functions and energies obtained from QCAD for a homojunction device with the same effective mass for the well and barrier. In this case, the wave functions and their first derivatives are all continuous across the junctions. The wave functions and energies agree very well with the corresponding analytic results in Figure 2.15, Ref.~\onlinecite{Harrison2005}. When the well and barrier have different effective masses, as shown in Fig.~\ref{fig:FiniteWFs}(b) for a heterojunction device, the wave functions are still continuous across the junctions, but their first derivatives are discontinuous due to the difference in effective masses.

\subsection{Self-Consistent Schrodinger-Poisson Solver}
\label{sec:SelfconSP}

In realistic quantum devices such as DQDs, we can divide the entire structure (relatively large) into semiclassical and quantum regions. These regions are chosen such that in semiclassical regions, solving the nonlinear Poisson equation alone is often sufficient to obtain a good estimate of electrostatics, whereas in quantum regions, the Poisson and Schrodinger equations need to be coupled self-consistently for electrons (we focus on electrons only as they are used for qubit operation). The coupled two equations take the following form
\begin{eqnarray}
\label{eq:CoupledPSEqns}
&& -\nabla \cdot (\epsilon_{s} \nabla \phi) = q [p(\phi) - n(E_i, \psi_i) + N_{D}^{+}(\phi) - N_{A}^{-}(\phi)], \nonumber \\
&& \frac{-\hbar^2}{2} \nabla \biggl( \frac{1}{m^*} \nabla \psi_i \biggl) + V(\phi, n) \psi_i = E_i \psi_i,
\end{eqnarray}
where the electron density $n$ becomes a function of the $i$th energy level $E_i$ and the envelope wave function $\psi_i$ of the Schrodinger equation, while the potential energy $V$ is a function of $\phi$ and $n$. The general expression for $n(E_i, \psi_i)$ is given by $\sum_{i} N_{i} |\psi_{i}|^{2}$, where the $N_{i}$ term takes different expressions depending on confinement dimensionality.

\subsubsection{Quantum Electron Density}
\label{sec:quantumedensity}

In Si quantum devices, when we focus on those devices where the Si/other material (e.g., Si/SiO$_2$) interfaces are parallel to the [100] plane, the six equivalent conduction band minima of the bulk silicon are split into two groups due to the breaking of crystallographic symmetry, widely known as $\Delta_4$ (fourfold degeneracy) and $\Delta_2$ (double degeneracy) valleys, with $\Delta_2$ valleys are lower in energy. At low temperatures, especially the operating temperatures for DQD qubits which are in the mK to a few Kelvin range, only the $\Delta_2$ valleys are occupied by electrons; therefore, we consider the $\Delta_2$ valleys only for Si devices in QCAD. Due to the ellipsoidal energy surfaces at the $\Delta_2$ minima, the electron effective mass in the Schrodinger equation is different from that used in computing the electron density, and it depends on the confinement direction and the number of confined directions.

\smallskip
\noindent \textbf{Two un-confined dimensions.}  In 1D-confined devices such as a 1D Si MOS capacitor, electrons are spatially confined in one direction (assumed $x$ direction in QCAD) but free to move in the $y$ and $z$ directions. The Si/SiO$_2$ interface is in the $y-z$ plane and perpendicular to the longitudinal axis of the $\Delta_2$ valleys. The coupled Schrodinger equation is 1D and given by
\begin{equation}
\label{eq:Schrodinger1D}
\frac{-\hbar^2}{2} \frac{d}{d x} \biggl( \frac{1}{m_l^*} \frac{d \psi_i(x)}{d x} \biggl) + V(\phi, n) \psi_i(x) = E_i \psi_i(x).
\end{equation}
where $m_l^*$ is the electron longitudinal effective mass of silicon. From Appendix A, the volume electron density $n$ is computed as
\begin{eqnarray}
\label{eq:VolEDens1DConfined}
n(E_i, \psi_i) &=& \sum_i N_i |\psi_i|^2 = \sum_i n_{2D,i} |\psi_i|^2 \nonumber \\
&=& \sum_i \biggl( 2 \frac{ m_t^* k_B T} {\pi \hbar^2} \ln \biggl[ 1 + \exp \biggl( \frac{E_F - E_i} {k_B T} \biggl) \biggl] |\psi_i|^2 \biggl), \nonumber \\
\end{eqnarray}
where $m_t^*$ is the electron transverse effective mass of silicon and 2 accounts for the double degeneracy of the $\Delta_2$ valleys. $|\psi_i(x)|^2$ is spatially normalized to 1, i.e., $\int |\psi_i(x)|^2 dx = 1$, and hence has the unit of 1/length. When $ E_F > E_i $, the $\exp[ (E_F-E_i)/(k_B T)]$ term in Eq.~(\ref{eq:VolEDens1DConfined}) can numerically go to infinity for very small T (mK to a few K), which can cause numerical instability. To avoid such problems, when the argument $(E_F - E_i)/(k_B T)$ is relatively large (e.g., $ > 100 $ ), we replace the $\ln [ 1 + \exp ( (E_F - E_i) / (k_B T) ) ]$ term with $(E_F - E_i) / (k_B T)$.

\smallskip
\noindent \textbf{One un-confined dimension.}  We next consider devices, such as quantum wire structures, where electrons are confined along two dimensions (assumed $x$ and $y$ directions in QCAD) and are free to move in the $z$ direction (the wire direction). The Si/SiO$_2$ interface is perpendicular to the $y$ axis and also the longitudinal axis of the $\Delta_2$ valleys. The coupled Schrodinger equation is 2D and given by
\begin{eqnarray}
\label{eq:Schrodinger2D}
&& \frac{-\hbar^2}{2} \frac{\partial}{\partial x} \biggl( \frac{1}{m_t^*} \frac{\partial \psi_i(x,y)}{\partial x} \biggl) - \frac{\hbar^2}{2} \frac{\partial}{\partial y} \biggl( \frac{1}{m_l^*} \frac{\partial \psi_i(x,y)}{\partial y} \biggl) \nonumber \\
&& + V(\phi, n) \psi_i(x,y) = E_i \psi_i(x,y).
\end{eqnarray}
From Appendix A, the volume electron density $n$ is computed as
\begin{eqnarray}
\label{eq:VolEDens2DConfined}
n(E_i, \psi_i) &=& \sum_i N_i |\psi_i|^2 = \sum_i n_{1D,i} |\psi_i|^2 \nonumber \\
&=& \sum_i \biggl[ 2 \biggl( \frac{2 m_t^* k_B T} {\pi \hbar^2} \biggl)^\frac{1}{2} \mathcal{F}_{-\frac{1}{2}} (\eta_F) \; |\psi_i|^2 \biggl],
\end{eqnarray}
where $\mathcal{F}_{-\frac{1}{2}} (\eta_F)$ is the Fermi-Dirac integral of -1/2 order. It is computed by using the approximate analytic expressions in Ref.~\onlinecite{Halen1985}, which have a very small error less than 0.001$\%$ in the entire $\eta_F$ range.  $|\psi_i(x,y)|^2$ is spatially normalized to 1, i.e., $\int \int |\psi_i(x,y)|^2 dx dy = 1$, and hence has the unit of 1/length$^2$.

\smallskip
\noindent \textbf{Zero un-confined dimensions.}  In devices such as quantum dot structures, electrons are spatially confined in all three directions and there are no (zero) dimensions in which they are free to move. The Si/SiO$_2$ interface is perpendicular to the $z$ direction and also the longitudinal axis of the $\Delta_2$ valleys. The coupled Schrodinger equation is 3D and given by
\begin{eqnarray}
\label{eq:Schrodinger3D}
&& \frac{-\hbar^2}{2} \frac{\partial}{\partial x} \biggl( \frac{1}{m_t^*} \frac{\partial \psi_i(x,y,z)}{\partial x} \biggl) - \frac{\hbar^2}{2} \frac{\partial}{\partial y} \biggl( \frac{1}{m_t^*} \frac{\partial \psi_i(x,y,z)}{\partial y} \biggl) \nonumber \\
&& - \frac{\hbar^2}{2} \frac{\partial}{\partial z} \biggl( \frac{1}{m_l^*} \frac{\partial \psi_i(x,y,z)}{\partial z} \biggl) \nonumber \\
&& + V(\phi, n) \psi_i(x,y,z) = E_i \psi_i(x,y,z).
\end{eqnarray}
The volume electron density $n$ is computed as
\begin{equation}
\label{eq:VolEDens3DConfined}
n(E_i, \psi_i) = \sum_i N_i |\psi_i|^2
= \sum_i \biggl[ \frac{4} {1+\exp(\frac{E_i - E_F} {k_B T})} \; |\psi_i|^2 \biggl],
\end{equation}
where 4 accounts for the double degeneracy of the $\Delta_2$ valleys and that of the spin. $\psi_i(x,y,z)$ is normalized to 1 in the 3D quantum domain, and has the unit of 1/length$^3$. When $ E_i > E_F $, the $\exp[ (E_i-E_F) / (k_B T) ]$ term in Eq.~(\ref{eq:VolEDens3DConfined}) can blow up numerically. To avoid such problem, when $ (E_i-E_F) / (k_B T) $ is relatively large (e.g., $> 100$ ), we replace the $ [1+\exp(\frac{E_i - E_F} {k_B T})]^{-1} $ term with $ \exp(\frac{E_F - E_i} {k_B T}) $.

All the above derivations are also applicable to other devices where the semiconductors have a single conduction band minimum located at the $\Gamma$ valley such as GaAs-based devices, except that the valley degeneracy is 1 and a single electron effective mass is used in all the equations.

Next we discuss the potential energy term $V(\phi, n)$ in the coupled Schrodinger equation. It takes the form of
\begin{equation}
\label{eq:SchrodPotEnergy}
V(\phi, n) = q \phi_{ref} - \chi - q\phi + V_{xc}(n),
\end{equation}
where $V_{xc}(n)$ is the exchange-correlation correction due to the Pauli exclusion principle in real many-electron systems. For the $V_{xc}(n)$ term, we use the well-known local density parameterization suggested by Hedin and Lundqvist \cite{Hedin1971} that has also been widely used by other authors \cite{Stern1984, VasileskaNanohub, Trellakis2004}. It is given as
\begin{eqnarray}
\label{eq:Vxc}
V_{xc}(n) &=& \frac{-q^2} {4 \pi^2 \epsilon_s} [3 \pi^2 n(\textbf{r})]^{\frac{1}{3}} \biggl[ 1+0.7734 x \ln \biggl( 1+\frac{1}{x} \biggl) \biggl], \nonumber \\
x &=& \frac{1}{21} \biggl( \frac{4 \pi n(\textbf{r}) b^3} {3}  \biggl)^{-\frac{1}{3}}, \nonumber \\
b &=& \frac{4 \pi \epsilon_s \hbar^2} {m_{xc}^* q^2}.
\end{eqnarray}
Since this parameterization requires a scalar effective mass $m_{xc}^*$ as input, we use an average mass for Si as suggested in Ref. \onlinecite{Trellakis2004},
\begin{equation}
\label{eq:mxc}
\frac{1}{m_{xc}^*} = \frac{1}{3} \biggl( \frac{1}{m_l^*} + \frac{2}{m_t^*} \biggl).
\end{equation}

\subsubsection{Self-Consistency}

The Schrodinger (S) and Poisson (P) equations in Eq.~(\ref{eq:CoupledPSEqns}) have strong nonlinear coupling. They need to be solved self-consistently by certain iterative numerical schemes. Various iteration schemes \cite{Stern1970, Moglestue1986, Laux1986, Kerkhoven1990, Sune1991, Luscombe1992, Trellakis1997, Trellakis2006} have been proposed and used over the past few decades. Among them, three are notable: the under-relaxation method, the damped Newton method, and the predictor-corrector approach.

The under-relaxation scheme \cite{Stern1970,Kerkhoven1990} (also called convergence-factor or simple average method) solves the S and P equations in succession, and under-relaxes the electron density $n$ or the electrostatic potential $\phi$ for the $k$th S-P outer iteration, using either a pre-set constant or an adaptively determined relaxation parameter $w^{(k)}$ (see the references for details). The advantage of this method is its simplicity. Its weakness is that the relaxation parameter $w^{(k)}$ is not known in advance and needs to be dynamically but heuristically readjusted during the course of iterations; if $w^{(k)}$ is too large, the iteration loop cannot reach convergence, whereas, if $w^{(k)}$ is too small, it takes too many iteration steps to achieve convergence.

The damped Newton method \cite{Laux1986, Luscombe1992} also solves the S and P equations in succession, but uses a damped Newton method \cite{Bank1980} for the outer S-P iteration. Specifically, this approach starts from an initial guess $\phi^{(0)}$, solves the Schrodinger eigenvalue problem, computes quantum electron density $n^{(k)}$ according to section \ref{sec:quantumedensity}, and then solves a linear P equation, obtained by linearizing the Poisson equation in Eq.~(\ref{eq:CoupledPSEqns}) according to the Newton method \cite{Bank1980} with an approximate Jacobian matrix; the potential $\phi_{out}^{(k)}$ from the linearized P equation is used to obtain $\phi_{in}^{(k+1)} = \phi_{in}^{(k)} + w^{(k)} (\phi_{out}^{(k)} - \phi_{in}^{(k)})$, which is then input to the S equation, and the procedure continues until self-consistency is reached. Here, the $w^{(k)}$ damping parameter is not heuristic, but can be determined by the selection algorithm of the damped Newton method (cf.~Ref.~\onlinecite{Bank1980}). $k$ represents the $k$th Newton iteration and also the $k$th outer S-P iteration. The Jacobian matrix must be approximated because the quantum electron density $n$ does not have explicit dependence on the potential $\phi$. The approximate Jacobian matrix is obtained by simply assuming a semi-classical electron density expression in the P equation. The approach has shown reasonably robustness \cite{Laux1986}, however, because of the approximate nature of the Jacobian, it often takes many Newton iterations (e.g., 50) to achieve sufficient self-consistent accuracy (e.g., $\phi$ is converged within 0.01 meV).

It is well-known that the under-relaxation \cite{Stern1970}, damped Newton \cite{Laux1986}, and other similar iteration schemes \cite{Moglestue1986, Sune1991} do not necessarily lead to convergence, or take too many iterations to achieve it. These schemes have been used mostly in 1D S-P problems and in only a limited number of 2D applications, and one may rightly expect that they would have much more difficulty in achieving convergence in 3D S-P problems (e.g.~in quantum dots). The key reason for the instability of these methods is that they do not physically address the strong nonlinear coupling between the S and P equations. In 1997 Trellakis et. al \cite{Trellakis1997} proposed the predictor-corrector (p-c) iteration scheme based on a perturbation argument. Due to its solid physical groundings, the p-c method has shown fast and robust convergence behavior \cite{Trellakis1997, Trellakis2000, Trellakis2006}, and has been widely used in 2D and 3D simulations of various quantum semiconductor devices \cite{Curatola2003, Khan2007, Wang2009}. Given its excellent track record, we implemented this p-c method in QCAD for the self-consistent S-P loop.

The key feature of the p-c method is that it partially decouples the S and P equations by moving most nonlinearities into the nonlinear Poisson equation
\begin{equation}
\label{eq:pcPoisson}
-\nabla \cdot (\epsilon_{s} \nabla \phi) = q [p(\phi) - \tilde{n}(\phi) + N_{D}^{+}(\phi) - N_{A}^{-}(\phi)],
\end{equation}
where $\tilde{n} (\phi)$ is an approximate expression for the quantum electron density $n(E_i, \psi_i)$, which has an explicit dependence on the potential $\phi$ (note the exact quantum density $n(E_i, \psi_i)$ does not have explicit dependence on $\phi$). The nonlinear Poisson equation can be solved by a Newton method (the predictor step). The predicted result for $\tilde{n}$ and $\phi$ from this equation is then corrected in an outer  iteration step by the solution of Schrodinger equation (the corrector step).

The approximate quantum density $\tilde{n}$ is obtained by using the first-order perturbation theory and the derivative property of Fermi-Dirac integrals \cite{Trellakis1997}. The resulting $\tilde{n}$ expression is the same as the exact quantum density $n$ given in Section \ref{sec:quantumedensity}, except that the argument in the Fermi-Dirac integral is modified to include an explicit dependence on $\phi$. For 1D-confined Si devices,  $\tilde{n}$ is given by
\begin{eqnarray}
\label{eq:ApproxEDens1DConfined}
\tilde{n}(\phi) &=& \sum_i \biggl( 2 \frac{ m_t^* k_B T} {\pi \hbar^2} |\psi_i^{(k)}|^2 \nonumber \\
&& \times \ln \biggl[ 1 + \exp \biggl( \frac{E_F - E_i^{(k)} + q(\phi - \phi^{(k)})} {k_B T} \biggl) \biggl] \biggl), \nonumber \\
\end{eqnarray}
where the superscripts $(k)$ denote quantities obtained in the previous $k$th outer S-P iteration step (hence they are known quantities). For 2D-confined Si devices,
\begin{eqnarray}
\label{eq:ApproxEDens2DConfined}
\tilde{n}(\phi) &=& \sum_i \biggl[ 2 \biggl( \frac{2 m_t^* k_B T} {\pi \hbar^2} \biggl)^\frac{1}{2} |\psi_i^{(k)}|^2 \nonumber \\
&& \times \mathcal{F}_{-\frac{1}{2}} \biggl( \frac{E_F - E_i^{(k)} + q (\phi - \phi^{(k)})} {k_B T}  \biggl) \biggl].
\end{eqnarray}
For 3D-confined Si devices,
\begin{eqnarray}
\label{eq:ApproxEDens3DConfined}
\tilde{n}(\phi) = \sum_i \biggl[ \frac{4 |\psi_i^{(k)}|^2 } {1 + \exp \biggl( \frac{E_i^{(k)} - E_F - q(\phi - \phi^{(k)})} {k_B T} \biggl) } \biggl].
\end{eqnarray}
Note that there is a minus sign in the $q(\phi - \phi^{(k)})$ term in Eq.~(\ref{eq:ApproxEDens3DConfined}). In principle, once the self-consistent S-P loop is converged, this term should be numerically zero, which might suggest that the sign shall not matter. However, our experience with QCAD is that the minus sign is very important for the 3D-confined case to achieve self-consistent convergence; if we used a plus sign here, the outer S-P loop ran into numerical oscillations.

The self-consistent p-c procedure in QCAD is done in the following steps.

(1) Solve the semiclassical nonlinear Poisson equation, Eq.~(\ref{eq:Poisson}), using the Newton solver in Trilinos \cite{Trilinos}, to obtain an initial potential $\phi^{(0)}$ and compute the initial total potential energy $V^{(0)}$ without the exchange-correlation correction $V_{xc}$.

(2) Solve the coupled Schrodinger equation for the $k$th ($k \ge 1$) S-P iteration step,
\begin{equation}
\frac{-\hbar^2}{2} \nabla \biggl( \frac{1}{m^*} \nabla \psi^{(k)} \biggl) + V^{(k-1)} \psi^{(k)} = E^{(k)} \psi^{(k)}, \nonumber
\end{equation}
to obtain $E^{(k)}$ and $\psi^{(k)}$ (performed using an eigensolver available in Trilinos).

(3) Solve the coupled nonlinear Poisson equation with the approximate quantum electron density $\tilde{n}^{(k)}(\phi^{(k)}; \phi^{(k-1)}, E^{(k)}, \psi^{(k)})$,
\begin{equation}
-\nabla \cdot (\epsilon_{s} \nabla \phi^{(k)}) = q [p(\phi^{(k)}) - \tilde{n}^{(k)} + N_{D}^{+}(\phi^{(k)}) - N_{A}^{-}(\phi^{(k)})], \nonumber
\end{equation}
using the Trilinos Newton solver to obtain the updated potential $\phi^{(k)}$, and compute $\tilde{n}^{(k)}$ and $V^{(k)}$ including $V_{xc}(\tilde{n}^{(k)})$. Note we want to use the latest electron density to compute $V_{xc}$ for good convergence.

(4) Check if $ || \phi^{(k)} - \phi^{(k-1)} || < \delta $ for $k \ge 2$ everywhere in the device, with $\delta$ being a pre-defined tolerance often chosen as $1\times10^{-5}$V; if not, repeat steps (2) to (4).

It is clear from the above procedure that there is no under-relaxation step between two S-P iterations and the outer iteration reduces to a simple alternation between solving S and P equations. In addition, the Newton Jacobian matrix for the nonlinear P equation, Eq.~(\ref{eq:pcPoisson}), can be found analytically, avoiding the necessity of using an approximate Jacobian matrix in the damped Newton iteration scheme \cite{Laux1986}. In terms of code implementation, the p-c method was very straightforward to implement in QCAD within the Albany framework.  Here we emphasize that, because of the automatic differential capability in Trilinos, the Newton Jacobian matrix is computed directly by the code, and we do not need to derive the Jacobian matrix. More details on the implementation and the Albany code structure are found in Ref. \onlinecite{Albany}.

\subsubsection{Validation Example}

\begin{figure}
\centering
\includegraphics[width=0.8\linewidth, bb=5 5 405 585]{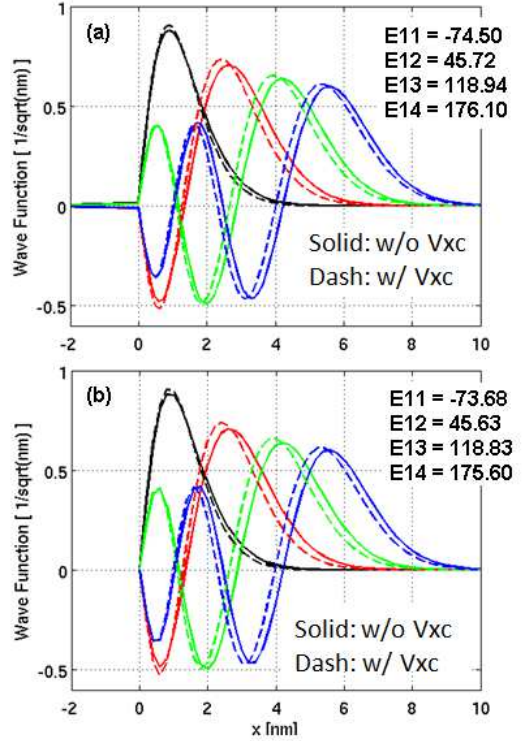}
\caption{\label{fig:1DPSWFs} (Color online) $\Delta_{2}$-valley lowest four subband wave functions and energies in a 1D MOS Si capacitor at T = 50 K and Vg = 3 V obtained from QCAD (a) and from SCHRED (b). The Si/SiO$_2$ interface is located at $x = 0$. The solid and dashed curves are obtained without and with the exchange-correlation effect, respectively. The subband energies in [meV], referenced from the Fermi level and including the $V_{xc}$ effect, are denoted by $\mathrm{E}1i$, where the ``1'' indicates the $\Delta_{2}$-valley and $i$ indexes the subband (SCHRED's labeling convention). For comparison, the corresponding energies without $V_{xc}$ are -72.54, 26.71, 90.73, 144.69 for QCAD, and -71.76, 26.12, 89.22, 142.27 for SCHRED. }
\end{figure}

To validate the self-consistent S-P solver, we performed simulations on two structures and compared with other simulation results. The first one is a 1D MOS Si capacitor with 4-nm oxide and $5\times 10^{17}$ cm$^{-3}$ p-substrate doping. Figure~\ref{fig:1DPSWFs} compares the $\Delta_{2}$-valley lowest four wave functions and energies in the capacitor obtained from QCAD and SCHRED \cite{Nanohub} at T = 50 K, the lowest temperature allowed by SCHRED. (SCHRED is a 1D self-consistent Poisson-Schrodinger solver for MOS capacitors available on www.nanohub.org.) There are two simulation differences between QCAD and SCHRED: (i) QCAD applies the S-P solver to both Si and SiO$_2$ regions, leading to slight wave function penetration in the oxide ($ x < 0$) as seen in Fig. \ref{fig:1DPSWFs}a, while SCHRED assumes $\psi = 0$ at the Si/SiO$_2$ interface; (ii) QCAD considers the two $\Delta_2$ valleys only, whereas SCHRED includes both the two $\Delta_2$ and the four $\Delta_4$ valleys. A typical effective mass of 0.5$m_0$ is often assumed \cite{Schneider1976, YPLi1985} for SiO$_2$, where $m_0$ is the free electron mass. To minimize the wave function difference near the Si/SiO$_2$ interface due to the different boundary conditions imposed by the two tools, we used 0.005$m_0$ as the SiO$_2$ effective mass for the QCAD simulations. The choice of setting $m_{ox}^{*} = 0.005 m_0$ is because, at the Si/SiO$_2$ interface, QCAD applies the flux conservation condition of $\frac{1} {m_{ox}^{*}} \cdot \frac{d \psi}{d x} |_{ox} = \frac{1} {m_{si}^{*}} \cdot \frac{d \psi}{d x} |_{si}$, and in order to make $\psi$ at the interface as close to 0 as possible (to be more consistent with SCHRED), we need small $\frac{d \psi}{d x} |_{ox}$, meaning small $m_{ox}^{*}$. At T = 50 K, we expect that ignoring the higher energy $\Delta_4$ valleys produces negligible effect on the $\Delta_2$-valley results, as only the $\Delta_2$-valley lowest subband is occupied by electrons at this low temperature. As expected, the wave functions and energies in Fig. \ref{fig:1DPSWFs} show excellent agreement between QCAD and SCHRED. The results also indicate that in this device, the exchange-correlation potential $V_{xc}$ significantly increases the subband energy separation due to the many-body interaction (e.g., the separation between the lowest two subbands is increased from 26.71+72.54 = 99 to 45.72+74.5 = 120 meV when including $V_{xc}$), and it also somewhat compresses the wave functions as seen by the differences between the dash and solid curves in the figure,  which agree with the observations in Ref. \onlinecite{VasileskaNanohub}.

\begin{figure}
\centering
\includegraphics[width=6.0cm, bb=120 310 440 560]{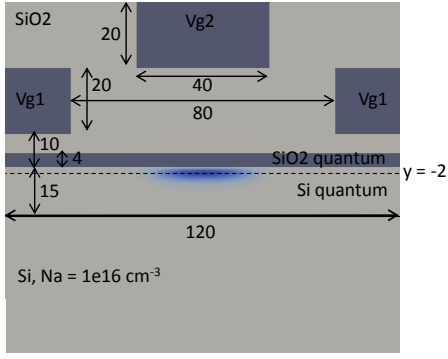}
\caption{\label{fig:SiQWire} (Color online) Schematic diagram of the simulated 2D structure with all dimensions given in nm. The blue 2D contour in the Si quantum region shows the $\Delta_2$-valley lowest subband wave function obtained from QCAD without the $V_{xc}$ effect at T = 10 K, $V_{g1}$ = 0.8 V, and $V_{g2}$ = 3.5 V with all voltages referred to flat band. The dash line denotes the $y = -2$ nm location. }
\end{figure}

\begin{figure}
\centering
\includegraphics[width=6.0cm, bb=105 235 490 545]{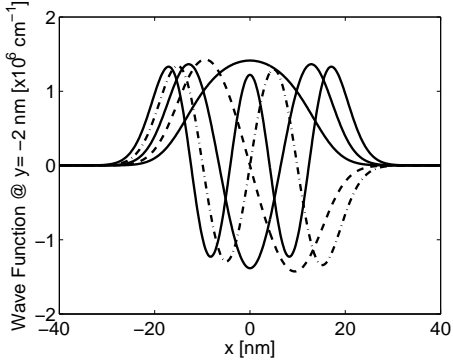}
\caption{\label{fig:SiQWire1DWFs} (Color online) $\Delta_2$-valley lowest five subband wave functions along the $y = -2$ nm dash line in Fig.~\ref{fig:SiQWire}. These wave functions agree very well with Fig. 4(a) in Ref.~\onlinecite{Laux1986APL}.}
\end{figure}

The second example is a gate-induced Si quantum wire structure from Ref.~\onlinecite{Laux1986APL}. Figure \ref{fig:SiQWire} shows the schematic diagram of the simulated 2D structure. For simulation purpose, the device is divided into quantum and semiclassical regions. The quantum regions include the 15-nm thick Si quantum and the 4-nm thick SiO$_2$ quantum regions denoted in the figure, where the self-consistent S-P solver is applied. The 15-nm and 4-nm were chosen such that the wave functions are essentially 0 at the boundaries of the quantum and non-quantum regions. The remaining Si and SiO$_2$ regions are treated as semiclassical, that is, only the Poisson equation with semiclassical carrier density is solved at each S-P iteration. The gate $V_{g2}$ induces electrons in the Si quantum region, while the $V_{g1}$ gates are used to deplete electrons, hence an effective quantum wire is formed with the wire direction perpendicular to the 2D plane. The blue 2D contour in the Si quantum region shows the $\Delta_2$-valley lowest subband wave function obtained from QCAD without the effect of $V_{xc}$ at T = 10 K, $V_{g1}$ = 0.8 V, and $V_{g2}$ = 3.5 V with all voltages referred to flat band. The peak of the wave function is located around the $y = -2$ nm dash line (the $y = 0$ location is at the Si-quantum/SiO$_2$-quantum interface). The lowest five subband wave functions along the $y = -2$ nm line are given in Fig.~\ref{fig:SiQWire1DWFs}, which agree very well with Fig. 4(a) in Ref.~\onlinecite{Laux1986APL}. Given T = 10 K and $V_{g1}$ = 0.8 V, we also performed QCAD S-P simulations for a range of $V_{g2}$ voltages, integrated the electron density in the Si quantum region for each $V_{g2}$, and then plotted the subband energies as a function of integrated electron density to compare with the results in Ref.~\onlinecite{Laux1986APL}. Figure~\ref{fig:SiQWireSubE} compares the $\Delta_2$-valley lowest three subband energies as a function of integrated electron density in the Si quantum region between QCAD and the reference. The agreement between them is very good considering the fact that the reference used a different S-P iteration scheme and did not mention if a fixed interface charge was used or not (no fixed charge was used in QCAD simulations).

\begin{figure}
\centering
\includegraphics[width=6.5cm, bb=95 235 490 535]{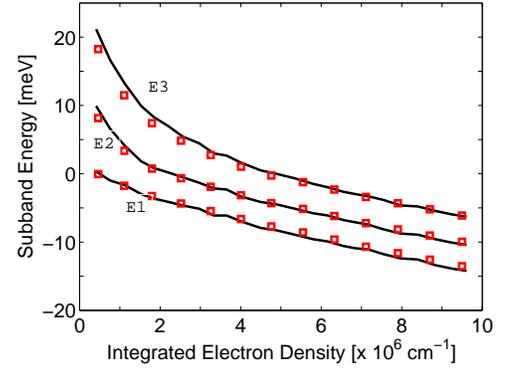}
\caption{\label{fig:SiQWireSubE} (Color online) $\Delta_2$-valley lowest three subband energies as a function of integrated electron density in the Si quantum region. Black curves plot the data extracted from Fig.~3 in Ref.~\onlinecite{Laux1986APL}, while the red squares are the data obtained from QCAD. The energies are with respect to the Fermi level which is set to 0. }
\end{figure}

\begin{figure}
\centering
\includegraphics[width=6.5cm, bb=95 235 480 550]{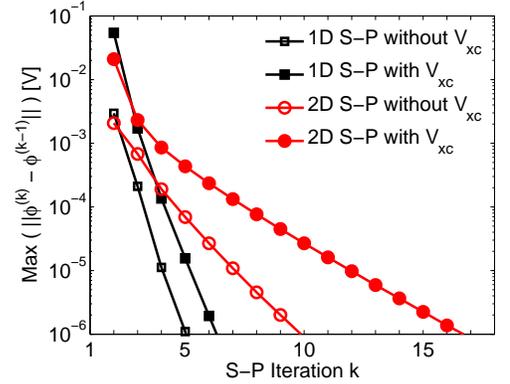}
\caption{\label{fig:SP1D2DConv} (Color online) Convergence behavior of the self-consistent QCAD S-P solver for the 1D MOS Si capacitor and the 2D gate-induced Si quantum wire devices. }
\end{figure}

For these 1D and 2D examples, their S-P convergence behavior from QCAD are plotted in Fig.~\ref{fig:SP1D2DConv}, where the vertical axis is the maximum potential error between two outer iterations in the entire device including the semiclassical region (note that the corresponding convergence are not available from either the referenced tool or paper). We see that the predictor-corrector approach for the S-P iteration leads to fast and monotonic convergence in the two closed quantum systems; and the inclusion of the $V_{xc}$ effect requires more iteration steps because of the stronger coupling that $V_{xc}$ introduces between the P and S equations.

\section{QD CAPACITANCE}
\label{sec:DQDCap}

In quantum dots, dot-to-gate capacitances are important quantities as they can be measured experimentally. Modeling of capacitances and comparison with experiment provide insight regarding the shape and location of a quantum dot and possible locations of defect charges. Capacitances can be computed in QCAD using either the semiclassical Poisson solver or the self-consistent S-P solver. In this section, we compare the capacitances found using both methods to those found experimentally.

\begin{figure}
\centering
\includegraphics[width=8.0cm, bb=45 195 505 675]{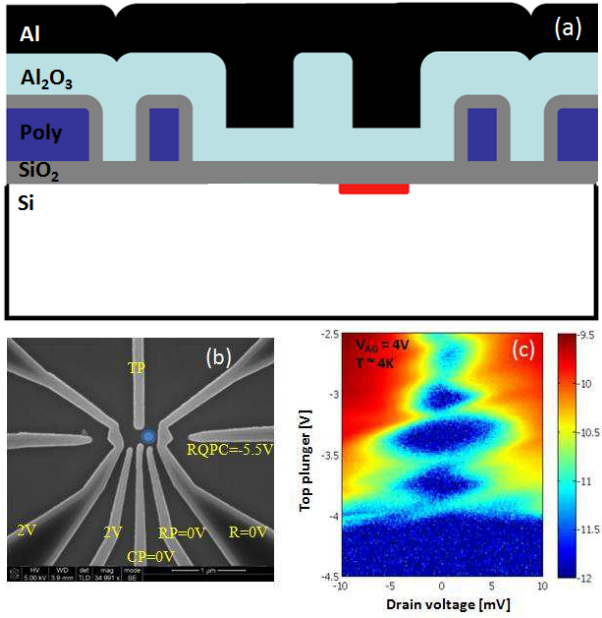}
\caption{\label{fig:SiDQDExperiment} (Color online) (a) Cross-section of an experimental Si DQD device, with the scanning electron micrograph (SEM) of the polysilicon depletion gates are shown in (b). The TP, CP, RP, R, and RQPC labels denote the depletion gates used to form the right dot. (c) Measured differential conductance for the right dot. }
\end{figure}

Figure~\ref{fig:SiDQDExperiment} shows the cross-section (a) and the SEM image of depletion gates (b) of an experimental Si DQD and the measured differential conductance for the right dot (c). The device is fabricated on a float-zone, high-resistivity (2-20 ohm-cm), p-type silicon substrate.  The depletion gates are made of polysilicon degenerately doped with arsenic.  A 60-nm Al$_2$O$_3$ insulation layer isolates a global Al top gate, which is positively biased to draw electrons into the quantum dot region. In the experiment, a quantum dot is formed on the right side of the nanostructure, schematically indicated as red square in (a) and as blue circle in (b).  The 2-V bias applied to the left gates induces 2DEG below them and fills those regions with electrons such that confinement of electrons only occurs on the right side. Further details about the fabrication can be found elsewhere \cite{Nordberg2009}.

The quantum dot is held at the temperature of liquid helium, approximately 4 K, and a standard lock-in technique is used to measure the differential conductance. The resulting conductance dependence on the top plunger gate bias as a function of the drain DC bias is shown in Fig.~\ref{fig:SiDQDExperiment}(c). The DC bias dependence of the Coulomb blockade through the quantum dot produces diamonds, typical of single dot behavior. The dot is observed to be empty after a last resonance around -4 V, indicative of the removal of the last electron. The total capacitance of the dot with only one electron is extracted from the charging energy, approximately 5 meV. Capacitances of the single-electron dot to other gates were extracted from the measured two-dimensional stability plots of the Coulomb blockade dependence on those gates in the device. The measured dot-to-gate capacitances are summarized in the experiment column of Table~\ref{table1}.

\begin{figure}
\centering
\includegraphics[width=4.0cm, bb=140 285 345 545]{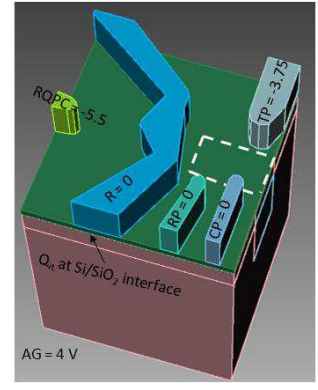}
\caption{\label{fig:SiDQDStruct} (Color online) Single dot structure with polysilicon depletion gates, gate oxide, and substrate shown for computing dot-to-gate capacitances. The TP, CP, RP, R, and RQPC labels correspond to those in Fig.~\ref{fig:SiDQDExperiment}(b) and their experimentally applied voltages are given in unit of [V]. AG is the voltage applied to the global Al top gate that is not shown. $Q_{it}$ is the fixed charge density at the Si/SiO$_2$ interface. The dashed white square denotes the top of the quantum dot region. }
\end{figure}

\begin{figure}
\centering
\includegraphics[width=4.2cm, bb=100 235 488 540]{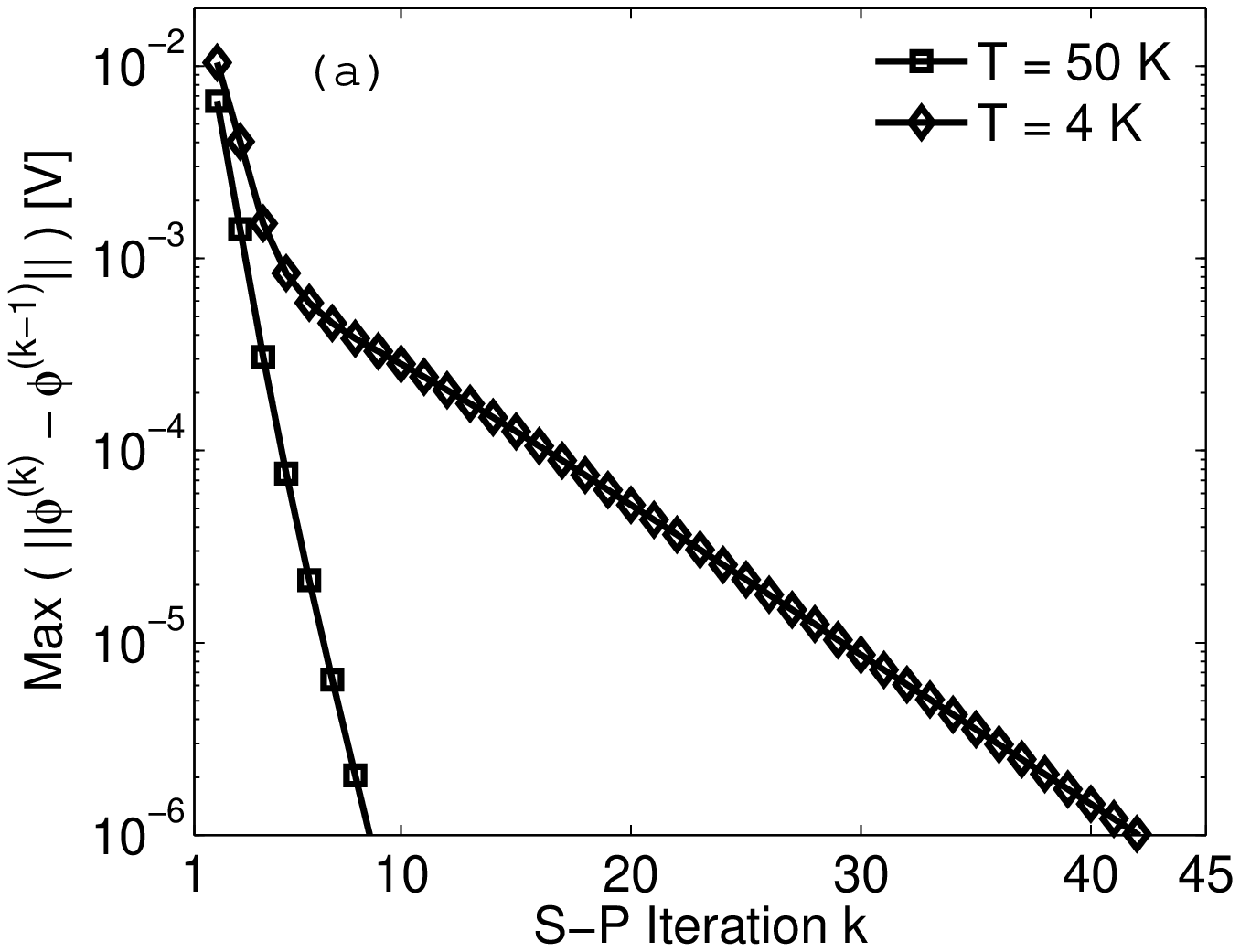}
\includegraphics[width=4.2cm, bb=100 235 488 540]{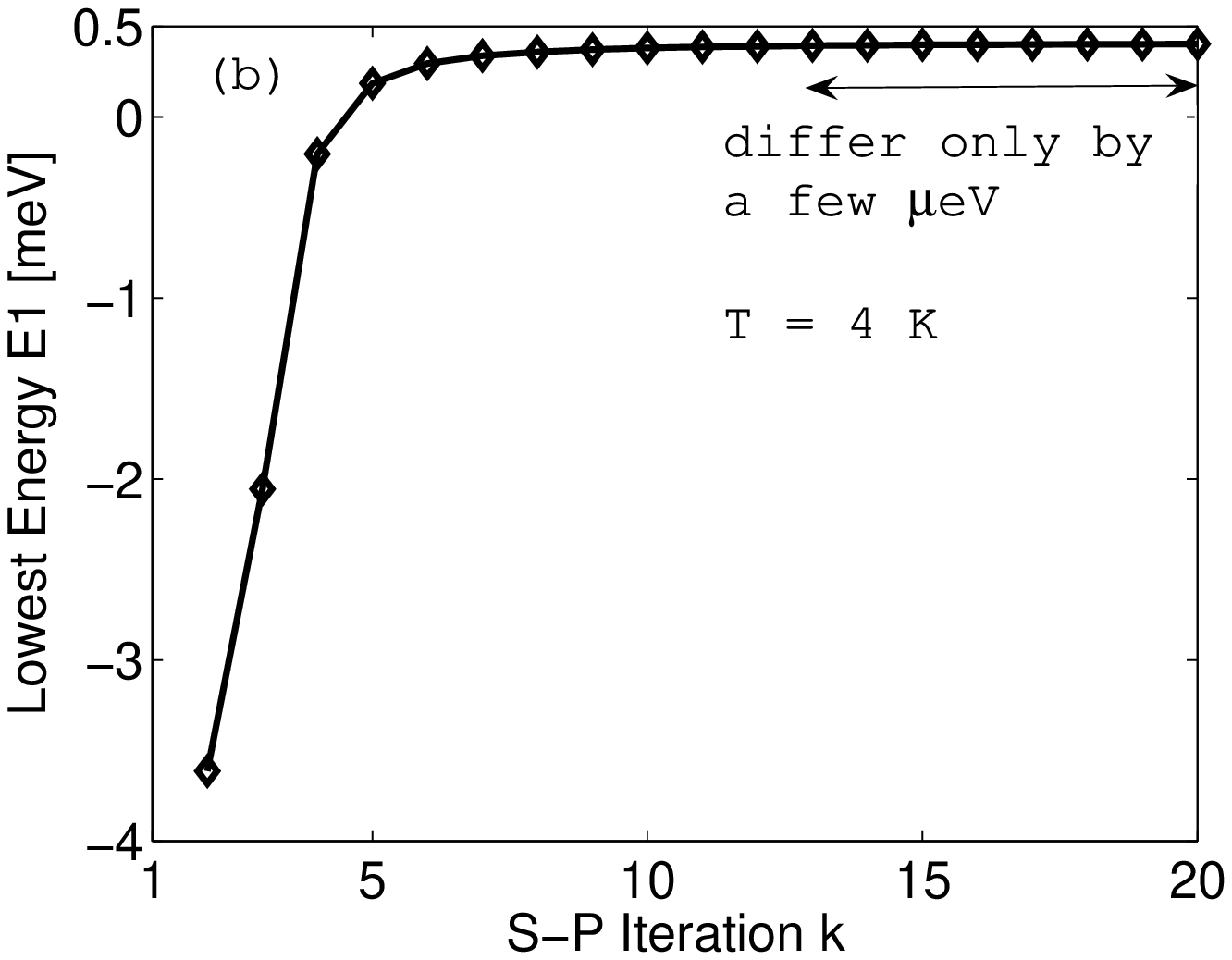}
\caption{\label{fig:SP3DConv} (a) Convergence behavior of the maximum electrostatic potential error of the self-consistent 3D QCAD S-P solver applied to the Si quantum dot. (b) Convergence behavior of the ground state energy E1 obtained by the S-P solver for the quantum dot at T = 4 K. }
\end{figure}

To simulate capacitances of the right dot to the various gates and compare them with experiment, we transferred the depletion gates SEM image to a CAD drawing and created a single dot structure that replicates the right dot of the experimental DQD device. The created structure is shown in Fig.~\ref{fig:SiDQDStruct} (for clarity, levels above the polysilicon depletion gates are not shown). The quantum region of this device is defined as a box volume of about 100-nm deep into the silicon substrate beneath the dashed white square. A positive voltage on the AG gate (not shown), which conformally covers the entire structure, induces a 2DEG in the silicon substrate under the Si/SiO$_2$ interface, while zero or negative voltages on the TP, CP, RP, and R gates deplete the 2DEG so that there is only one electron in the quantum dot. The RQPC gate is used for modulating a charge-sensing channel which can be used to measure the charge in the dot experimentally. Since the real device always contains an unknown number of defect charges in many locations, our QCAD simulations use an effective charge density $Q_{it}$ at the Si/SiO$_2$ interface as a tuning parameter. $Q_{it}$ is chosen such that the integrated electron density within the quantum region is equal to one under the given experimental voltages. Lumping all the possible effects of disorders into a single tuning parameter is clearly simplistic, but it avoids a time-consuming calibration process and gives a crude approximation of the real conditions while allowing for relatively fast calculation and analysis of the capacitances.

To compute the dot-to-gate capacitances including quantum effects, we apply the self-consistent 3D S-P solver to simulate the quantum region, while using the semiclassical Poisson solver for the rest of the device at each S-P iteration step. We see in Fig.~\ref{fig:SP1D2DConv} that the predictor-corrector scheme for the S-P loop leads to monotonic convergence for the 1D and 2D devices examined. For the 3D quantum dot, the S-P solver even with the $V_{xc}$ effect also shows monotonic convergence as shown in Fig.~\ref{fig:SP3DConv}(a) at lattice temperatures of 50 K and 4 K. Although the convergence of the potential is relatively slow at 4 K (42 iterations are needed to reach a maximum potential error of less than $1\times10^{-6}$V), the ground state energy E1 converges within much fewer iterations. As shown in Fig.~\ref{fig:SP3DConv}(b), the energy E1 reaches an error of a few $\mu$eV as early as the 13th iteration. In practice, we find that sufficient accuracy is obtained by running the simulation for 15 to 20 iterations at T = 4 K. For even lower temperatures such as in the milli-Kelvin range, more iteration steps may be required.

\begin{table}
  \caption{\; Simulated and measured dot-to-gate capacitances for the single quantum dot shown in Fig.~\ref{fig:SiDQDStruct} operated in the one-electron regime.}
  \centering
  \begin{tabular}{| l | l | l | l | l | l |}
    \hline
          & Exper & QCAD  & QCAD & QCAD & QCAD \\
          & iment & Poisson & 3D S-P & Poisson & 3D S-P \\
     \hline
    T [K] & 4 & 50 & 50 & 4 & 4 \\
    \hline
    Q$_{it}$ [$\times$ $10^{11}$ & unknown & -4.61 & -4.51 & -4.54 & -4.43 \\
    cm$^{-2}$] & & & & & \\
    \hline
    e$^{-}$ in the dot & 1 & 0.96 & 0.96 & 0.95 & 0.95 \\
    \hline
    dot-AG [aF] & 2.37 & 3.98 & 4.33 & 4.38 & 4.98 \\
    \hline
    dot-TP [aF] & 0.48 & 0.33 & 0.37 & 0.37 & 0.44 \\
    \hline
    dot-CP [aF] & 0.54 & 0.86 & 0.96 & 0.95 & 1.12 \\
    \hline
    dot-RP [aF] & 0.29 & 0.64 & 0.72 & 0.71 & 0.84 \\
    \hline
    dot-R [aF]  & 0.56 & 2.07 & 2.30 & 2.28 & 2.64 \\
    \hline
  \end{tabular}
  \label{table1}
\end{table}

The calculated dot-to-gate capacitances are compared with experiment in Table~\ref{table1}. The capacitance is computed using the static form of
\begin{eqnarray}
C &=& \frac{dQ} {dV} = \frac{q(N_2 - N_1)} {dV}, \nonumber \\
N_{1,2} &=& \int \int \int n_{1,2}(x,y,z) dx dy dz, \nonumber
\end{eqnarray}
where $N_{1,2}$ is the number of electrons in the quantum region for two given voltages with small difference $dV$ ($dV$ is fixed at 0.01 V) applied to the same gate. The QCAD Poisson simulations at 50 K were carried out to compare with the semiclassical results (not shown) obtained in the past by using the commercial Sentaurus Device tool \cite{SDevice2010} which could not converge at temperatures below 50 K. The semiclassical capacitances at 50 K obtained by Sentaurus are similar to those by the QCAD Poisson solver as expected. From the table, we see that, while the simulated capacitances do not match the measurement exactly, they have the correct order of magnitude, and the relative magnitude of values within one column correlates reasonably well with the other columns. We expect that the discrepancy between the simulated and measured values is due to the over-simplified treatment of defect charges in the simulations. In particular, the higher simulated dot-to-depletion-gate (CP, RP, and R) capacitances indicate that, there could be defect charges associated with these gates that need to be taken into account.

We observe that, for a given temperature and with a constraint of one electron in the dot, the capacitances calculated by the self-consistent 3D S-P solver are somewhat higher than the semiclassical values by the Poisson solver. One often supposes that quantum confinement will reduce capacitance, which is true if the number of electrons is allowed to change. In Table~\ref{table1}, the electron number is fixed at approximately one in the dot. To understand the higher quantum capacitances (i.e., capacitances obtained by using the QCAD 3D S-P quantum solver), we performed analytical analysis on the 50-K case and also at 4 K. The electron density in the quantum dot region at T = 50 K is shown in Fig.~\ref{fig:SiDQDDens}. The conduction band 1D profiles along the depth up to 10-nm deep denoted by the white lines are plotted in Fig.~\ref{fig:SiDQDCB1D}. The choice of 10-nm depth is because electron density is small below the depth and can be neglected in the analysis.  The 1D profiles are approximated by linear fits and used for analytical calculations. The profile obtained from the QCAD Poisson solver is well approximated by the linear fit, while the profile from the 3D S-P solver deviates from a linear fit near the Si/SiO$_2$ interface due to the inclusion of quantum confinement. The linear fits produce electric field of $7.045 \times 10^6$ V/m for the Poisson case and $8.256 \times 10^6$ V/m for the S-P case. Linear fits to the 1D band profiles allow for analytical calculations of semiclassical and quantum electron densities in triangular wells, which are described in Appendix B.

\begin{figure}
\centering
\includegraphics[width=7.0cm, bb=65 290 475 490]{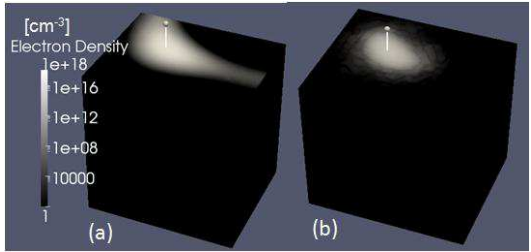}
\caption{\label{fig:SiDQDDens} Electron density in the quantum dot region of the device shown in Fig.~\ref{fig:SiDQDStruct} at T = 50 K obtained from the QCAD semiclassical Poisson solver (a), and from the self-consistent 3D S-P solver (b). The white vertical lines denote the locations of 1D cuts along the z direction.}
\end{figure}

The analytical electron densities for the extracted 1D triangular wells are compared with the 1D cuts from the QCAD 3D solvers in Fig.~\ref{fig:SiDQDeDens1D}. The agreement between the QCAD 1D cuts and the analytical solutions is exceptionally good, which indicates that the lateral ($x$-$y$ plane) confinement is weak and the vertical ($z$ direction) confinement dominates the solution in the Si quantum dot. The weak contribution of the lateral confinement can be seen from the ground state energy $E_1$, which is equal to 9.79 meV (with respect to $E_F = 0$) from the QCAD 3D S-P solver, whereas it is 7.11 meV (the 25 meV shift is already included) from the analytic solution given in Appendix B. The slightly higher $E_1$ value from QCAD can be attributed to the lateral weak confinement. The very good agreement between QCAD and analytic indicates that we can use the analytical expressions to study how the capacitance varies between semiclassical and quantum calculations. From Appendix B, we know that the derivative of integrated electron density with respect to electric field, i.e., $d \int q n(z) dz / d F$ represents a capacitance per unit length. As shown in Fig.~\ref{fig:SiDQDCap1D}, the analytical quantum solution shows higher capacitance in the field range of interest, which is consistent with the observation in Table~\ref{table1}. Intuitively, this is because given a fixed number of electrons in the dot, the quantum electron density is shifted away from the interface and much broader in space (see Fig.~\ref{fig:SiDQDeDens1D}), which allows for somewhat higher charge variation for a given voltage change, leading to higher capacitance. Similar analysis and results also hold at T = 4 K.

\begin{figure}
\centering
\includegraphics[width=7.0cm, bb=70 300 475 620]{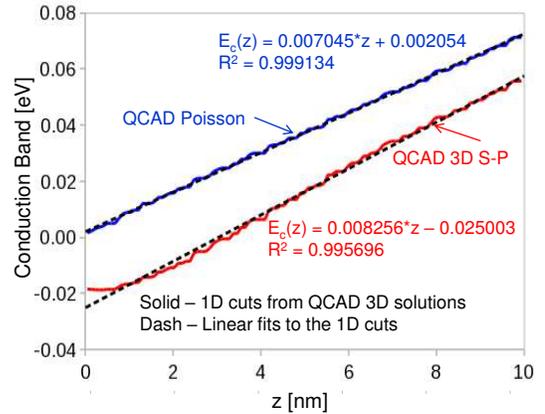}
\caption{\label{fig:SiDQDCB1D} (Color online) Conduction band profiles as a function of depth inside Si substrate starting from the Si/SiO$_2$ interface. The locations of the 1D cuts are indicated by the vertical lines in Fig.~\ref{fig:SiDQDDens}. Solid lines are the 1D cuts from QCAD 3D solutions, while dash lines are linear fits to the 1D cuts. The little wiggles on the solid lines are due to interpolation noise. The Fermi level $E_F$ is set to 0.}
\end{figure}

\begin{figure}
\centering
\includegraphics[width=7.0cm, bb=70 215 510 575]{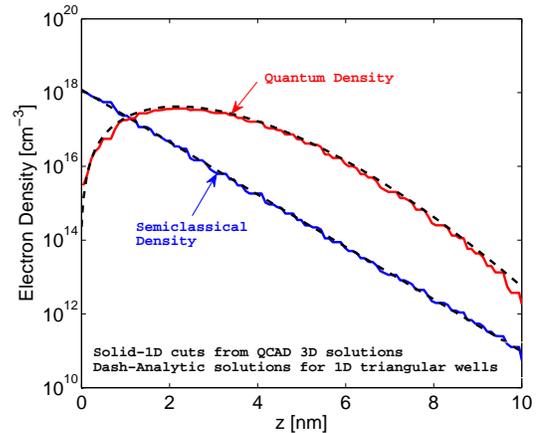}
\caption{\label{fig:SiDQDeDens1D} (Color online) Electron density comparison between 1D cuts from the QCAD 3D solvers and the analytical solutions for the 1D triangular wells obtained in Fig.~\ref{fig:SiDQDCB1D}. The agreement between the QCAD 1D cuts and the analytical solutions is exceptionally good, which indicates that the lateral ($x$-$y$ plane) confinement is weak and the vertical ($z$ direction) confinement dominates the solution in the Si quantum dot.}
\end{figure}

\begin{figure}
\centering
\includegraphics[width=7.0cm, bb=70 205 510 560]{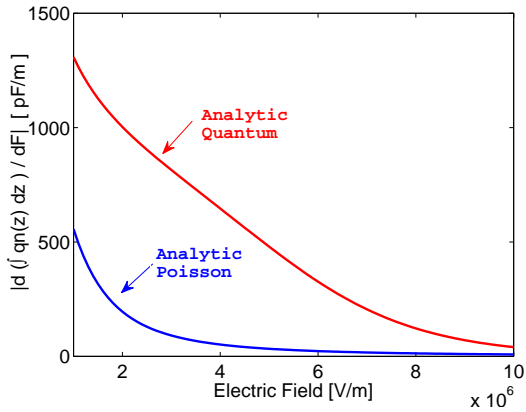}
\caption{\label{fig:SiDQDCap1D} (Color online) Analytical derivative of integrated electron density with respect to electric field in 1D triangular wells. The derivative has the unit of pF/m, representing a capacitance per unit length. The field range is selected such that it contains the fields extracted in Fig.~\ref{fig:SiDQDCB1D} which keep the same number of electrons between Poisson and quantum calculations. For the entire field range, the analytical quantum solution shows higher capacitance, which is consistent with the observation in Table~\ref{table1}. }
\end{figure}

\section{DQD OPTIMIZATION}
\label{sec:DQDOpt}

The ability to tune a DQD device so that there are few electrons in its quantum dots while its tunnel barriers are controllable is very important for experimentally manipulating and measuring the device as a qubit. However, it is often challenging to achieve this, and the search for ``good'' DQD designs is difficult and time-consuming, due to the large space of possible depletion gate layouts and the many choices related to which materials are used and the thickness of the various device layers. The ability to rapidly search over this large design space was one of the primary motivations for developing QCAD, and in this section we give an example of how the QCAD tool can be used to address the design of double quantum dot qubits.

Figure~\ref{fig:DepletionGates} shows the top view of four typical depletion gate patterns (transferred from SEM images) used in experimental DQD devices. For a given depletion gate layout, it often takes several months to find out experimentally if the device can work in the few-electron regime or not.  Using QCAD's optimization capabilities (provided by Dakota), we can perform optimization simulations of many device designs simultaneously on parallel clusters and can quickly determine whether a device is a few-electron DQD candidate or not. Each individual optimization would allow specified gate voltages to vary as adjustable parameters in order to meet one or multiple targets that are believed to be true of ``good'' DQD qubit designs. Such targets include (i) a given number of electrons in the left and/or right dot, e.g., one electron in the left dot in Fig.~\ref{fig:DepletionGates}(a); (ii) specified values for the electron densities at multiple tunnel barriers, e.g., LTB, DB, and LQPCB in Fig.~\ref{fig:DepletionGates}(a), which indicate whether the tunnel barriers can be turned on/off; (iii) the distance between a charge sensing constriction and a dot.

\begin{figure}
\centering
\includegraphics[width=8.0cm, bb=40 210 490 560]{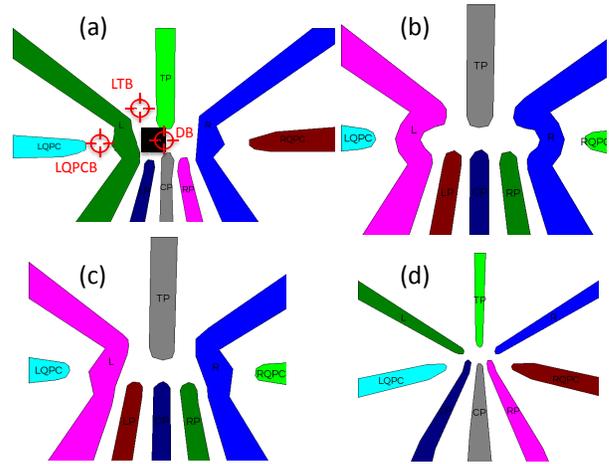}
\caption{\label{fig:DepletionGates} (Color online) Top view of exemplary depletion gate patterns (transferred from SEM images) in experimental DQD devices. Each finger in the figures indicates a metal or polysilicon gate that can be set to a different voltage to form quantum dots. The red circles in (a) define the approximate locations for the left tunnel barrier (LTB), the dot barrier (DB), and the left QPC barrier (LQPCB). The black square in (a) denotes the left dot region.}
\end{figure}

The locations of tunnel barriers in QCAD are dynamically detected through a saddle point searching algorithm based on the Nudged Elastic Band approach \cite{Henkelman2000}.  This algorithm loosely solves the classical dynamics of a discretized elastic band stretched between two points.  After the elastic band relaxes, the highest point on the band marks the saddle point. We have adopted several enhancements to the traditional nudged elastic band, including a climbing force added to the current highest point of the band (making it the ``Climbing nudged elastic band'' method) and a penalty term which discourages the band from making sharp turns weighted by an ``anti-kinking'' factor (a Gaussian smooth function).  The initial and final points of the elastic band are determined either by direct user input or by maximizing a quantity of interest within a specified region.  The latter case is used when we desire one end of the band to lie on a quantum dot whose location is a priori unknown: the starting (or ending) point is specified as the point of maximum electron density within the region where we expect the quantum dot to form.

Figure~\ref{fig:SiDQDOptDens}(a) shows a top view of the electron density at the Si/SiO$_2$ interface for the device shown in Fig.~\ref{fig:DepletionGates}(a) after optimization. In this optimization, the depletion gates (TP, CP, LP tied to RP, L tied to R, LQPC tied to RQPC) and the top Al gate AG (not shown) voltages are allowed to vary, and the targets are to obtain one electron (integrated electron density close to one) in the left dot and simultaneously keep the electron densities in the LTB, DB, and LQPCB barriers at specified values (e.g., $10^{18}$ cm$^{-3}$ at all three barriers). The gate voltages that meet these optimization targets are given in the caption of Fig.~\ref{fig:SiDQDOptDens}(a). For this device, the optimization results suggest that it can possibly operate in the one/few-electron regime with all tunnel barriers controllable. Experimental results have corroborated this observation, as measurements on this device have shown good few-electron characteristics. Optimization results for a different structure, shown in Fig.~\ref{fig:SiDQDOptDens}(b), suggest a different situation where the RQPCB barrier is completely shut off even when there are about 100 electrons in each dot. In this case, QCAD's optimal solution could not meet its given targets simultaneously -- one could find solutions for the target of having one electron in each dot, \emph{or} for the target of setting electron densities to given values at the tunnel barriers, but not both, indicating that this device may not be able to reach the few-electron regime. This was also confirmed experimentally, as devices of this geometry did not display good few-electron behavior. These results show that the ability to perform device optimizations can help to efficiently narrow down device candidates that meet essential requirements.

\begin{figure}
\centering
\includegraphics[width=7.6cm, bb=60 485 460 640]{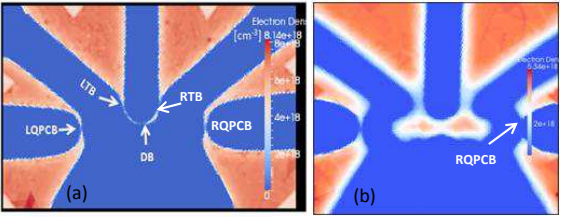}
\caption{\label{fig:SiDQDOptDens} (Color online) (a) View along the Si/SiO$_2$ interface of the resulting electron density after the optimizing the depletion gate voltages to obtain one electron in the left dot while keeping the electron densities in the LTB, DB, and LQPCB barriers at $10^{18}$ cm$^{-3}$.  The device used is the same as that shown in Fig.~\ref{fig:DepletionGates}(a) (T = 0.2 K). The gate voltages that meet the optimization targets are AG = 3.48 V, TP = -0.74 V, CP = -0.007 V, LP = RP = -5.95 V, L = R = -1.15 V, and LQPC = RQPC = -2.41 V. (b) Top view of the electron density after optimization for a different DQD device, where there are about 100 electrons in each dot. }
\end{figure}

We have run optimizations such as the one described above on dozens of DQD device designs (some already fabricated, others not). Overall, these optimizations have helped us answer three DQD design questions that are critical in achieving few-electron QD behavior: (i) which devices allow one electron in each dot and also simultaneous control of tunnel barriers; (ii) for a given device, do tunnel barriers turn on before or after a dot has many electrons; (iii) what are the locations and shapes of the main dots and charge sensing constrictions.

\section{CONCLUSION}
\label{sec:Conclusion}

We have developed a versatile finite-element-based tool called QCAD that is aimed at efficient simulation of multi-dimensional quantum devices, in particular, semiconductor multi-QD devices for use as qubits. QCAD is able to solve for the electrostatic potential with or without quantum effects (e.g. quantum confinement), and self-consistent single- and/or multi-electron wave functions and energies in 1D/2D/3D quantum devices. We have demonstrated robust convergence of the tool even at near-zero-Kelvin temperatures.  The self-consistent quantum models in QCAD allow for analysis of quantum effects on device parameters of importance, and we have shown this for the case of dot-to-gate capacitances. Very high QCAD simulation throughput is achieved through pre- and post-processing scripting, distributed parallel computing capability and resources. QCAD simulations and optimizations of realistic multi-QDs allow for fast and valuable design comparison, optimization, and guidance to accelerate the experimental development of few-electron multi-QD qubits.

\begin{acknowledgments}
This work was supported by the Laboratory Directed Research and Development (LDRD) program at Sandia National Laboratories. Sandia is a multiprogram laboratory operated by Sandia Corporation, a Lockheed Martin Company, for the United States Department of Energy's National Nuclear Security Administration under Contract DE-AC04-94AL85000.
\end{acknowledgments}

\appendix

\section{}
All derivations here assume single conduction band minimum and isotropic effective mass. In bulk devices (no spatial confinement), electrons are free to move in all three directions. Denote $\textbf{r} = (x, y, z)$ and $\textbf{k} = (k_x, k_y, k_z)$, and assume parabolic energy dispersion $E = E_C + \hbar^2 k^2 / (2 m^*)$ with $k = |\textbf{k}|$. The 3D electron density in bulk devices is given as
\begin{equation}
n_{3D} = \frac{2}{\Omega} \sum_{\textbf{k}} f(E_F - E)  \nonumber,
\end{equation}
where 2 includes spin degeneracy and $f(E)$ is the Fermi-Dirac distribution, i.e., $f(E) = [1+ \exp(\frac{-E}{k_B T})] ^{-1}$. Making use of the sum-to-integral rule,
\begin{equation}
\frac{1}{\Omega} \sum_\textbf{k} = \frac{1}{\Omega} \frac{\Omega}{(2 \pi)^3} \int d^3 \textbf{k} \nonumber,
\end{equation}
and converting the integrand to spherical coordinates as the energy $E$ only depends on the amplitude of $\textbf{k}$,
\begin{equation}
\frac{1}{\Omega} \sum_\textbf{k} = \frac{1}{(2 \pi)^3} \int d^3 \textbf{k} = \frac{1}{(2 \pi)^3} \int_0^{+\infty} 4 \pi k^2 dk \nonumber,
\end{equation}
then the electron density becomes
\begin{equation}
n_{3D} = \frac{1}{\pi^2} \int_0^{+\infty} \frac{k^2 dk} {1 + \exp( \frac{E - E_F} {k_B T} )}. \nonumber
\end{equation}
Since $E = E_C + \hbar^2 k^2 / (2 m^*)$, we obtain
\begin{equation}
k^2 dk = \frac{1}{2} \biggl( \frac{2 m^*} {\hbar^2} \biggl) ^{\frac{3}{2}} \sqrt{E- E_C} \; \; dE, \nonumber
\end{equation}
then
\begin{equation}
n_{3D} = \frac{1}{2 \pi^2} \int_{E_C}^{+\infty} \biggl( \frac{2 m^*} {\hbar^2} \biggl) ^{\frac{3}{2}} \frac{\sqrt{E- E_C}} {1 + \exp( \frac{E - E_F} {k_B T} )} \; dE. \nonumber
\end{equation}
From the above expression, we can define the density of states (DOS) for bulk semiconductors, that is,
\begin{equation}
\label{eq:DOS3D}
g_{3D}(E) = \frac{1}{2 \pi^2} \biggl( \frac{2 m^*} {\hbar^2} \biggl) ^{\frac{3}{2}} \sqrt{E- E_C} \nonumber.
\end{equation}
With this definition, $n_{3D}$ can be rewritten as
\begin{equation}
n_{3D} = \int_{E_C}^{+\infty} g_{3D}(E) f(E_F - E) dE, \nonumber
\end{equation}
which is the familiar expression widely found in solid state textbooks. Let $\varepsilon = (E - E_C) / (k_B T) $ and $ \eta_F = (E_F - E_C) / (k_B T)$, then
\begin{eqnarray}
\label{eq:EDens3D}
n_{3D} &=& \frac{1}{2 \pi^2} \biggl( \frac{2 m^* k_B T} {\hbar^2} \biggl) ^{\frac{3}{2}} \int_0^{+\infty} \frac{\sqrt{\varepsilon} d \varepsilon } {1+\exp(\varepsilon - \eta_F)}, \nonumber \\
&=& 2 \biggl( \frac{m^* k_B T} {2 \pi \hbar^2} \biggl) ^{\frac{3}{2}} \frac{2}{\sqrt{\pi}} \int_0^{+\infty} \frac{\sqrt{\varepsilon} d \varepsilon } {1+\exp(\varepsilon - \eta_F)}, \nonumber \\
&=& N_C \mathcal{F}_{1/2} (\eta_F),
\end{eqnarray}
where $N_C$ is defined as the effective DOS in the conduction band and $\mathcal{F}_{1/2}(\eta_F)$ is the Fermi-Dirac integral of 1/2 order \cite{Blakemore1982}.

In 1D-confined devices such as quantum well structures, electrons are free to move in two directions, whereas the third direction is spatially confined and forms quantized energy levels. Assume the free directions are along $y$ and $z$, and the confined direction is along $x$. Denote $\textbf{r} = (y, z)$ and $\textbf{k} = (k_y, k_z)$. The parabolic energy dispersion becomes $E = E_i + \hbar^2 k^2 / (2 m^*)$, where $E_i$ is the $i$th quantized energy level determined by the 1D Schrodinger equation in the confined direction $x$. The 2D electron density in the $y-z$ plane for a given $E_i$ is defined as
\begin{equation}
n_{2D,i} = \frac{2}{A} \sum_\textbf{k} f(E_F - E). \nonumber
\end{equation}
Following a procedure similar to the bulk case, we have
\begin{eqnarray}
\label{eq:EDens2D}
n_{2D,i} &=& \frac{2}{A} \frac{A}{(2 \pi)^2} \int d^2 \textbf{k} f(E_F - E) \nonumber \\
&=& \frac{2}{(2 \pi)^2} \int_0^{+\infty} 2 \pi k dk f(E_F - E) \nonumber \\
&=& \frac{1}{\pi} \int_0^{+\infty} \frac{k dk} {1 + \exp ( \frac{E - E_F}{k_B T} ) } \nonumber \\
&=& \frac{m^*} {\pi \hbar^2} \int_{E_i}^{+\infty} \frac{dE} {1 + \exp(\frac{E-E_F}{k_B T}) } \nonumber \\
&=& \frac{m^* k_B T} {\pi \hbar^2} \ln \biggl[ 1 + \exp \biggl( \frac{E_F - E_i} {k_B T} \biggl) \biggl] \nonumber \\
&=& \frac{m^* k_B T} {\pi \hbar^2} \mathcal{F}_0 (\eta_F),
\end{eqnarray}
where $\mathcal{F}_0 (\eta_F)$ is the Fermi-Dirac integral of zero order with $\eta_F = (E_F - E_i) / (k_B T)$ here. In the final step of the derivation, the integral identity
\begin{equation}
\int \frac{dy} {1+e^y} = \int \frac{e^{-y} dy} {1+e^{-y}} = - \ln(1+e^{-y}) \nonumber
\end{equation}
is used. It is worthy of noting that one can obtain the 2D DOS from Eq.~(\ref{eq:EDens2D}), $g_{2D}(E) = m^* / (\pi \hbar^2)$, which is independent of energy.

In 2D-confined devices such as quantum wire structures, electrons are free to move in only one direction, whereas the other two directions are spatially confined and form quantized energy levels. Assume the free direction is along $z$, and the confined directions are along $x$ and $y$. The parabolic energy dispersion becomes $E = E_i + \hbar^2 k_z^2 / (2 m^*)$, where $E_i$ is the $i$th quantized energy level determined by the 2D Schrodinger equation in the confined directions. The 1D electron density in the $z$ direction for a given $E_i$ is defined as
\begin{equation}
n_{1D,i} = \frac{2}{L} \sum_{k_z} f(E_F - E). \nonumber
\end{equation}
Following a procedure similar to the bulk case, we have
\begin{eqnarray}
n_{1D,i} &=& \frac{2}{L} \frac{L}{2 \pi} \int_{-\infty}^{+\infty} d k_z f(E_F -E) \nonumber \\
&=& \frac{2}{\pi} \int_0^{+\infty} d k_z f(E_F -E) \nonumber \\
&=& \frac{1}{\pi} \biggl( \frac{2 m^*} {\hbar^2} \biggl)^\frac{1}{2} \int_{E_i}^{+\infty} \frac{(E - E_i)^{-\frac{1}{2}} \; dE} {1 + \exp(\frac{E- E_F} {k_B T} ) } \; . \nonumber
\end{eqnarray}
From this expression, we can define the 1D DOS as follows,
\begin{equation}
\label{eq:DOS1D}
g_{1D}(E) = \frac{1}{\pi} \biggl( \frac{2 m^*} {\hbar^2} \biggl)^\frac{1}{2} (E - E_i)^{-\frac{1}{2}}. \nonumber
\end{equation}
Let $\varepsilon = (E- E_i) / (k_B T)$ and $\eta_F = (E_F - E_i) / (k_B T)$, then
\begin{eqnarray}
\label{eq:EDens1D}
n_{1D,i} &=& \frac{1}{\pi} \biggl( \frac{2 m^* k_B T} {\hbar^2} \biggl)^\frac{1}{2} \int_0^{+\infty} \frac{\varepsilon^{-\frac{1}{2}} \; d \varepsilon} {1 + \exp(\varepsilon - \eta_F) } \nonumber \\
&=& \biggl( \frac{2 m^* k_B T} {\pi \hbar^2} \biggl)^\frac{1}{2} \frac{1}{\sqrt{\pi}} \int_0^{+\infty} \frac{\varepsilon^{-\frac{1}{2}} \; d \varepsilon} {1 + \exp(\varepsilon - \eta_F) } \nonumber \\
&=& \biggl( \frac{2 m^* k_B T} {\pi \hbar^2} \biggl)^\frac{1}{2} \mathcal{F}_{-\frac{1}{2}} (\eta_F),
\end{eqnarray}
where $\mathcal{F}_{-\frac{1}{2}} (\eta_F)$ is the Fermi-Dirac integral of -1/2 order \cite{Halen1985}.

\section{}

Given the extracted 1D triangular well for the QCAD Poisson case in Fig.~\ref{fig:SiDQDCB1D}, $E_C(z) = 0.007045 z + 0.002054 $ eV $= q F z + b$ with the electric field $F = 7.045 \times 10^6 $ V/m and the intercept $b = 0.002054$ eV, the semiclassical 3D electron density can be calculated as (assuming MB statistics for simplicity),
\begin{equation}
\label{eq:SemiDensTriWell}
n(z) = N_C \exp \biggl( \frac{E_F - E_C} {k_B T} \biggl)= N_C \exp \biggl( \frac{E_F - qFz - b} {k_B T} \biggl), \nonumber
\end{equation}
with $N_C$ given in Eq.~(\ref{eq:EffectiveDOS}). Integrating the electron density along $z$ leads to
\begin{equation}
\label{eq:IntSemiDensTriWell}
\int_0^{z_f} n(z) dz = \frac{N_C k_B T} {q F} \exp \biggl( \frac{E_F - b} {k_B T} \biggl) \biggl[ \exp \biggl( \frac{-q F z_f} {k_B T} \biggl) -1 \biggl], \nonumber
\end{equation}
which corresponds to a sheet electron density per unit area. $z_f$ is an upper bound in $z$ above which the electron density is negligible and it is set to 10 nm in correspondence with Fig.~\ref{fig:SiDQDCB1D}. Taking derivative of this expression with respect to the field gives
\begin{eqnarray}
\label{eq:SemiDerivOverF}
\frac{d \int_0^{z_f} q n(z) dz} {d F} &=& q N_C \exp \biggl( \frac{E_F - b} {k_B T} \biggl)  \nonumber \\
&& \times \biggl[ \biggl( \frac{k_B T} {q F^2} + \frac{z_f} {F} \biggl)  \exp \biggl( \frac{-q F z_f} {k_B T} \biggl) - \frac{k_B T}{q F^2}  \bigg], \nonumber \\
\end{eqnarray}
which represents a capacitance per unit length.

Given the extracted 1D triangular well for the QCAD S-P case in Fig.~\ref{fig:SiDQDCB1D}, $E_C(z) = 0.008256 z - 0.025003 $ eV, we first shift the profile up to obtain $E_C(z) = 0.008256 z$ eV $ = q F z $ with $F = 8.256 \times 10^6$ V/m, which leads to analytic solutions for the wave functions and eigen-energies, and then we shift the eigen-energies down by the same amount to obtain the actual energies. The $i$th analytic wave function for the 1D triangular well given by $E_C(z) = q F z $ takes the form of \cite{Soln4TriWell}
\begin{eqnarray}
\label{eq:WFTriWell}
\psi_i(z) &=& A \; Ai[u(z)], \nonumber \\
u(z) &=& \bigg( \frac{2 m_l^{*} q F} {\hbar^2} \bigg)^{\frac{1}{3}} \biggl(z - \frac{E_i}{q F} \biggl), \nonumber \\
E_i &=& \biggl( \frac{\hbar^2 q^2 F^2} {2 m_l^*} \biggl)^{\frac{1}{3}} \biggl[ \frac{3 \pi}{2} \biggl(i - \frac{1}{4} \biggl)  \biggl]^{\frac{2}{3}}, \nonumber \\
A &=& \biggl( \frac{2 m_l^* q F} {\hbar^2} \biggl)^{\frac{1}{6}} \biggl[ Ai^{\prime 2} (\lambda_0) - \lambda_0 Ai^2(\lambda_0) \biggl]^{\frac{-1}{2}}, \nonumber \\
\lambda_0 &=& -\frac{E_i} {q F} \biggl( \frac{2 m_l^* q F} {\hbar^2} \biggl)^{\frac{1}{3}}, i = 1, 2, ...
\end{eqnarray}
where $A$ is the normalization factor, $m_l^*$ is the longitudinal effective mass of Si, and $Ai(\lambda_0)$ is the Airy function. At T = 50 K and lower temperatures, and because of the low electron density in the DQD device, only the lowest subband is occupied, hence the volume quantum electron density can be computed as
\begin{equation}
\label{eq:QuanDensTriWell}
n(z) = \frac{2 m_t^* k_B T} {\pi \hbar^2} \ln \biggl[ 1 + \exp \biggl( \frac{E_F - E_1} {k_B T} \biggl) \biggl] |\psi_1(z)|^2, \nonumber
\end{equation}
with 2 accounting for the $\Delta_2$-valley double degeneracy. $E_1$ here is equal to the $E_1$ in Eq.~(\ref{eq:WFTriWell}) minus 0.025003 eV. As the wave function modulus is normalized, i.e., $\int_0^{+\infty} |\psi_1(z)|^2 d z = 1 $, integrating the electron density leads to
\begin{equation}
\int_0^{+\infty} n(z) dz = \frac{2 m_t^* k_B T} {\pi \hbar^2} \ln \biggl[ 1 + \exp \biggl( \frac{E_F - E_1} {k_B T} \biggl) \biggl]. \nonumber
\end{equation}
The derivative of the sheet electron density with respect to the field is given by
\begin{eqnarray}
\label{eq:QuanDerivOverF}
\frac{d \int_0^{+\infty} q n(z) dz} {d F} &=& \frac{2 m_t^* k_B T} {\pi \hbar^2} \biggl[ 1 +  \exp \biggl( \frac{E_1 - E_F} {k_B T} \biggl) \biggl]^{-1} \nonumber \\
&& \times \frac{-1}{k_B T} \frac{d E_1}{d F}, \nonumber \\
\frac{d E_1}{d F} &=& \biggl( \frac{3 \pi^2 \hbar^2 q^2} {16 m_t^* F} \biggl)^{\frac{1}{3}},
\end{eqnarray}
which also represents a capacitance per unit length.


\begin{thebibliography}{10}

\bibitem{Morello2010} A. Morello, J. J. Pla, et al., Nature {\bf 467}, 687 (2010).

\bibitem{Borselli2011} M. G. Borselli, K. Eng, et al., Appl. Phys. Lett. {\bf 99}, 063109 (2011).

\bibitem{Lim2011} W. H. Lim, C. H. Yang, F. A. Zwanenburg, and A. S. Dzurak, Nanotechnology {\bf 22}, 335704 (2011).

\bibitem{Yamahata2012} G. Yamahata, T. Kodera, H. O. H. Churchill, K. Uchida, C. M. Marcus, and S. Oda, Phys. Rev. B {\bf 86}, 115322 (2012).

\bibitem{Tracy2010} L. A. Tracy, E. P. Nordberg, et al., Appl. Phys. Lett. {\bf 97}, 192110 (2010).

\bibitem{TMLu2011} T. M. Lu, N. C. Bishop, et al., Appl. Phys. Lett. {\bf 99}, 043101 (2011).

\bibitem{Laird2010} E. A. Laird, J. M. Taylor, D. P. DiVincenzo, C. M. Marcus, M. P. Hanson, and A. C. Gossard, Phys. Rev. B {\bf 82}, 075403 (2010).

\bibitem{Koh2012} T. S. Koh, J. K. Gamble, M. Friesen, M. A. Eriksson, and S. N. Coppersmith, Phys. Rev. Lett. {\bf 109}, 250503 (2012).

%\bibitem{Rahman2012} R. Rahman, E. Nielsen, R. P. Muller, and M. S. Carroll, Phys. Rev. B {\bf 85}, 125423 (2012).

\bibitem{Stopa1996} M. Stopa, Phys. Rev. B {\bf 54}, 13767 (1996).

\bibitem{Milicic2000} S. N. Milicic, F. Badrieh, D. Vasileska, A. Gunther, and S. M. Goodnick, Superlatt. and Microstruct. {\bf 27}, 377 (2000).

\bibitem{Friesen2003} M. Friesen, P. Rugheimer, D. E. Savage, M. G. Lagally, D. W. Weide, R. Joynt, and M. A. Eriksson, Phys. Rev. B {\bf 67}, 121301 (2003).

\bibitem{Melnikov2005} D. V. Melnikov, J. Kim, L. X. Zhang, and J. P. Leburton, IEE Proc.-Circuits Devices Syst. {\bf 152}, 377 (2005).

\bibitem{Klimeck2008} B. Muralidharan, H. Ryu, Z. Huang, and G. Klimeck, J. Comput. Electron. {\bf 7}, 403 (2008).

\bibitem{SDevice2010} \emph{Sentauru Device User Guide}, Version D-2010.03, Synopsys Inc. and http://www.quantumwise.com/.

\bibitem{Nanohub} http://www.nanohub.org/.

\bibitem{Trilinos} http://trilinos.sandia.gov, http://dakota.sandia.gov, and http://cubit.sandia.gov.

\bibitem{Albany} A. G. Salinger, R. P. Pawlowski, et al., ACM Trans. Math. Software, submitted (2013).

\bibitem{Gao2012} X. Gao, E. Nielsen, R. P. Muller, R. W. Young, A. G. Salinger, N. C. Bishop, and M. S. Carroll, Proc. of 15th IWCE, DOI: 10.1109/IWCE.2012.6242832 (2012).

\bibitem{NielsenCI2010} E. Nielsen and R. P. Muller, arXiv:1006.2735 (2010).

\bibitem{Anderson_2009} Christopher R. Anderson, J. Comp. Phys. {\bf 228}, 4745 (2009).

\bibitem{Mohan1995} N. Mohankumar and A. Natarajan, Phys. Stat. Sol.(b) {\bf 188}, 635 (1995).

\bibitem{Blakemore1982} J. S. Blakemore, Solid-State Electron. {\bf 25}, 1067 (1982).

\bibitem{PolyLogarithm} See for example http://mathworld.wolfram.com/Polylogarithm.html

\bibitem{Kim2008} R. Kim and M. Lunstrom, arXiv:0811.0116 (2008).

\bibitem{Press2007} W. H. Press, S. A. Teukolsky, W. T. Vetterling, and B. P. Flannery, \emph{Numerical Recipes: The Art of Scientific Computing} (3rd Ed., Cambridge University Press, 2007).

\bibitem{Bedn1978} D. Bednarczyk and J. Bednarczyk, Phys. Lett. A {\bf 64}, 409 (1979).

\bibitem{Halen1985} P. Van Halen and D. L. Pulfrey, J. Appl. Phys. {\bf 57}, 5271 (1985); ibid, J. Appl. Phys. {\bf 59}, 2264 (1986).

\bibitem{VasileskaNanohub} D. Vasileska, \emph{Solving the Effective Mass Schrodinger Equation in State-of-the-Art Devices}, http://nanohub.org.

\bibitem{Lundstrom1981} M. S. Lundstrom, R. J. Schwartz, and J. L. Gray, Solid-Stat. Electron. {\bf 24}, 195 (1981).

\bibitem{Tan_SchroPo_1990} I‐H. Tan, G. L. Snider, L. D. Chang, and E. L. Hu, J. Appl. Phys. {\bf 68}, 4071 (1990).

\bibitem{SzeBook} S. M. Sze and Kwok K. Ng, \emph{Physics of Semiconductor Devices} (3rd Ed., John Wiley and Sons Ltd., 2007).

\bibitem{Trellakis2000} A. Trellakis and U. Ravaioli, Comput. Methods Appl. Mech. Engrg. {\bf 181}, 437 (2000).

\bibitem{Nilsson1973} N. G. Nilsson, Phys. Stat. Solidi.(a) {\bf 19}, K75 (1973).

\bibitem{RobinBCs} Ronald B. Guenther and John W. Lee, \emph{Partial Differential Equations of Mathematical Physics and Integral Equations} (Dover Publications Inc., 1996).

\bibitem{Harrison2005} P. Harrison, \emph{Quantum Wells, Wires and Dots} (2nd Ed., John Wiley and Sons Ltd., 2005).

\bibitem{Hedin1971} L. Hedin and B. I. Lundqvist, J. Phys. C: Solid State Phys. {\bf 4}, 2064 (1971).

\bibitem{Stern1984} F. Stern and S. Das Sarma, Phys. Rev. B {\bf 30}, 840 (1984).

\bibitem{Trellakis2004} A. Trellakis and U. Ravaioli, Solid State Electronics {\bf 48}, 367 (2004).

\bibitem{Stern1970} F. Stern, J. Comput. Phys. {\bf 6}, 56 (1970).

\bibitem{Moglestue1986} C. Moglestue, J. Appl. Phys. {\bf 59}, 3175 (1986).

\bibitem{Laux1986} S. E. Laux and A. C. Warren, IEDM Technical Digest, 567 (1986); S. E. Laux and F. Stern, Appl. Phys. Lett. {\bf 49}, 91 (1986).

\bibitem{Kerkhoven1990} T. Kerkhoven, A. G. Galick, U. Ravaioli, J. H. Arends, and Y. Saad, J. Appl. Phys. {\bf 68}, 3461 (1990).

\bibitem{Sune1991} J. Sune, P. Olivo, and B. Ricco, J. Appl. Phys. {\bf 70}, 337 (1991).

\bibitem{Luscombe1992} J. H. Luscombem, A. M. Bouchard, and M. Luban, Phys. Rev. B {\bf 46}, 10262 (1992).

\bibitem{Trellakis1997} A. Trellakis, A. T. Galick, A. Pacelli, and U. Ravaioli, J. Appl. Phys. {\bf 81}, 7880 (1997).

% \bibitem{Trellakis1998} A. Trellakis, A. T. Galick, A. Pacelli and U. Ravaioli, VLSI DESIGN {\bf 8}, 105 (1998).

\bibitem{Trellakis2006} A. Trellakis, T. Andlauer, and P. Vogl, \emph{Large Scale Scientific Computing} (Springer-Verlag Berlin Heidelberg, 2006).

\bibitem{Bank1980} R. E. Bank and D. J. Rose, SIAM J. Numer. Anal. {\bf 17}, 806 (1980).

\bibitem{Curatola2003} G. Curatola, G. Iannaccone, Comput. Mater. Sci. {\bf 28}, 342 (2003).

\bibitem{Khan2007} H. R. Khan, D. Mamaluy, and D. Vasileska, IEEE Trans. Electron. Devices {\bf 54}, 784 (2007).

\bibitem{Wang2009} H. Wang, G. Wang, S. Chang, and Q. Huang, Micro \& Nano Lett. {\bf 4}, 122 (2009).

\bibitem{Schneider1976} P. M. Schneider and W. B. Fowler, Phys. Rev. Lett. {\bf 36}, 425 (1976).

\bibitem{YPLi1985} Y. P. Li and W. Y. Ching, Phys. Rev. B {\bf 31}, 2172 (1985).

\bibitem{Laux1986APL} S. E. Laux and F. Stern, Appl. Phys. Lett. {\bf 49}, 91 (1986).

\bibitem{Nordberg2009} E. P. Nordberg, G. A. Ten Eyck, et al., Phys. Rev. B {\bf 80}, 115331 (2009).

\bibitem{Henkelman2000} G. Henkelman, B. P. Uberuaga, and H. Jonsson, J. Chem. Phys. {\bf 113}, 9901
(2000).

\bibitem{Soln4TriWell} \url{http://www.iue.tuwien.ac.at/phd/gehring/node116.html}, and \url{http://ecee.colorado.edu/~bart/book/book/chapter1/ch1_2.htm}.


\end{thebibliography}
\end{document}